\documentclass[onecolumn,floatfix,superscriptaddress,a4paper,
               showpacs,showkeys,nofootinbib,preprint]{revtex4}

\usepackage{epsfig}
\usepackage{amssymb,latexsym,amsmath}
\newcommand{\eq}[1]{\begin{align} #1 \end{align}}

\begin{document}

\title{Power Law in Micro-Canonical Ensemble with\\ Scaling Volume Fluctuations}
%\bf  Power Law in Statistical Model}

 \author{V. V. Begun}
 \affiliation{Bogolyubov Institute for Theoretical Physics, Kiev, Ukraine}

 \author{M. Ga\'zdzicki}
 \affiliation{Institut f\"ur Kernphysik, University of Frankfurt, Frankfurt, Germany}
 \affiliation{Jan Kochanowski University, Kielce, Poland}

 \author{M. I. Gorenstein}
 \affiliation{Bogolyubov Institute for Theoretical Physics, Kiev, Ukraine}
 \affiliation{Frankfurt Institute for Advanced Studies, Frankfurt,Germany}

%-----------------------------------------------------------
\begin{abstract}
Volume fluctuations are introduced in a statistical modelling of
relativistic particle collisions. The micro-canonical ensemble is
used, and the volume fluctuations are assumed to have the specific
scaling properties. This leads to the KNO scaling of the particle
multiplicity distributions as measured in p+p interactions. A
striking prediction of the model is a power law form of the single
particle momentum spectrum at high momenta. Moreover, the mean
multiplicity of heavy particles also decreases as a function of
the particle mass according to a power law. Finally, it is shown
that the dependence of the momentum spectrum on the particle mass
and momentum reduces to the dependence on the particle energy.
These results resemble the properties of particle production in
collisions of high energy particles.

\end{abstract}



\maketitle

%-------------------------------------------------

\section{ Introduction}

In  collisions at relativistic energies many new particles are
produced. Their number, masses and charges as well as their
momenta vary from event to event. Most of the experimental results
concern  single particle production properties averaged over many
interactions. It is well established that some of these
properties, namely, mean particle multiplicities and transverse
momentum spectra, follow simple rules of statistical mechanics.
In proton-proton (p+p) collisions
%\footnote{Similar results are
%observed in $p+\overline{p}$ and $e^++e^-$ collisions.}
the single particle momentum distribution  has an approximately
Boltzmann form \cite{Ha:65} in the local rest frame of produced
matter:
%\footnote{ Bose and
%Fermi statistics for mesons and baryons respectively, lead only to
%statistical mechanics parameters extracted from fitting the
%data.}:
\begin{equation}\label{exp}
\frac{dN}{p^2dp}~\sim~
\exp\left(-~\frac{\sqrt{p^2+m^2}}{T}\right)~,
\end{equation}
where $T$, $p$ and $m$  are the temperature parameter,  the
particle momentum and its mass, respectively. At large momentum,
$p\gg m$, Eq.~(\ref{exp}) gives:
\eq{\label{exp1}
\frac{dN}{p^2dp}~\sim~ \exp\left(-~\frac{p}{T}\right)~.
}
Integration of (\ref{exp}) over momentum yields the mean particle
multiplicity, $\langle N \rangle$, which is also governed by the
Boltzmann factor for $m>>T$:
\begin{equation}\label{exp2}
\langle N \rangle ~ \sim~ (mT)^{3/2}~ \exp\left(-~\frac{m}{T}\right)~.
\end{equation}
The approximate validity of the exponential distributions
(\ref{exp}-\ref{exp2}) is confirmed by numerous experimental
results on bulk particle production in high energy collisions. The
agreement is limited to the low transverse momentum ($p_T \leq 2$
GeV) and the low mass ($m \leq 2$ GeV) domains. However, the
temperature parameter $T$ extracted from the data on p+p
interactions is in the range 160-190~MeV~\cite{Be:97}. Thus,
almost all particles are produced at low $p_T$ and with low
masses.

\vspace{0.2 cm} Along with evident successes there are obvious
problems of the statistical approach. The probability $P(N)$ to
create $N$ particles in p+p collisions obeys the KNO scaling
\cite{kno} (see also Refs.~\cite{knog1,knog2,knog3}), namely:
\begin{equation}\label{kno}
 P(N)~ =~ \langle N \rangle^{~-1}~\psi(z)~,
\end{equation}
where $\langle N \rangle$ is the mean multiplicity and the KNO
scaling function $\psi(z)$ only depends on $z\equiv N/\langle
N\rangle $. The mean multiplicity increases with increasing
collision energy, whereas the KNO scaling function remains
unchanged. The latter implies that the scaled variance $\omega$ of
the multiplicity distribution $P(N)$ grows linearly with the mean
multiplicity:
\begin{equation}\label{kno1}
 \omega~ \equiv~
\frac{\langle N^2 \rangle~-~\langle N \rangle^2 } {\langle N
\rangle}~\propto ~\langle N \rangle~.
\end{equation}
A qualitatively different behavior is predicted
within the existing statistical models
\cite{Be:04,mce1,Be:07}, namely the scaled variance is expected to
be independent of the mean multiplicity:
\begin{equation}\label{kno2}
 \omega ~\approx~ const~\approx 1~.
\end{equation}
This contradiction between the data and the statistical models
constitutes the first problem which will be considered in this
paper.

%-------------------------------------------------
The second and the third problems  which will be addressed here
concern particle production at high (transverse) momenta and with
high masses, respectively. In these regions the single particle
energy distribution seems to obey a power law
behavior~\cite{powerlaw}:
\begin{equation}\label{P}
\frac{dN}{p^2dp}~\sim~ \left(\sqrt{p^2+m^2}\right)^{-K}~.
\end{equation}
At $p\gg m$, Eq.~(\ref{P}) gives:
\begin{equation}\label{P1}
\frac{dN}{p^2dp}~ \sim ~p^{-K_p}~,
\end{equation}
with $K_p=K$. Integration of (\ref{P}) over particle momentum
yields the mean multiplicity which follows a power law dependence
on the particle mass:
\begin{equation}\label{P2}
\langle N \rangle~ \sim~ ~m^{-K_m}~,
\end{equation}
with $K_m=K_p-3$. The Eqs.~(\ref{P1}) and (\ref{P2}) approximately
describe the data on spectra of light particles at large ($p\geq
3$~GeV) (transverse) momenta and on the mean multiplicity of heavy
($m\geq 3$~GeV) particles, respectively. The power law parameters
fitted to the data  are $K_p\cong 8$ and $K_m\approx
5$~\cite{powerlaw}. \noindent One observes a growing disagreement
between the exponential behavior (\ref{exp1}), (\ref{exp2}) and
power law dependence (\ref{P1}), (\ref{P2}) with increasing
(transverse) momentum and mass. At $p=10$~GeV or $m =10$~GeV the
statistical models underestimate the data by more than 10 orders
of magnitude.
%On the other hand, the same functional dependence of
%the transverse momentum spectra and mean multiplicities of heavy
%particles, and the relation between the power parameters,
%$K_m=K_p-3$, describing these spectra, were derived in
%Ref.~\cite{powerlaw} within the framework of the statistical
%approach. This was done by replacing the Boltzmann single particle
%energy distribution by the power law dependence. The above results
%motivated
%the effort to extend validity of the statistical approach to the .

%\vspace{0.2 cm} {\bf A power law at very high (transverse) momenta
%is described by perturbative QCD, however the intermediate
%momentum region $p>5$ GeV is not accessible neither by statistical
%models nor by pQCD.}

%\vspace{0.2 cm}
%However, the
%QCD predictions are not accessible for the hadron production at
%low transverse momenta and low masses (soft domain). Thus,
%{\bf Nowadays two different approaches are used to describe
%particle production: the statistical hadron model for the soft
%domain (low transverse momenta and low hadron masses) and the
%perturbative QCD for the hard domain.
%An extension of the QCD
%to the soft domain requires further development of the
%non-perturbative methods and calculations.
%
%\vspace{0.2 cm}
In the present paper we make an attempt to extend the statistical
model to the hard domain of high transverse momenta and/or high
hadron masses (hard domain).
% which provides a possible solution of the
%problems mentioned above.
The proposal is inspired by statistical
type regularities \cite{powerlaw} in the high transverse mass
region, as well as by the recent work on the statistical ensembles
with fluctuating extensive quantities \cite{alpha}. We postulate
that the volume of the system created in p+p collision changes
from event to event\footnote{ The volume fluctuations in hadron
statistical physics were first introduced in the framework of the
isobaric ensemble in Refs.~\cite{pressure,stas}.}. The main
assumptions of the proposed approach are the following.
\begin{enumerate}
\item Each final state created in p+p interactions is identified
with a micro-state of a micro-canonical ensemble (MCE) defined by
the volume $V$, energy $E$, and conserved charges of the system.
By definition of the MCE, all its micro-states appear with the
same probability. \item The volume of micro-canonical ensembles
represented in p+p interactions fluctuates from collision to
collision. The volume probability density function,
$P_{\alpha}(V)$, is given by the scaling function, $P_{\alpha}(V)
= \overline{V}^{-1} \phi_{\alpha}(V/\overline{V})$, where
$\overline{V}$ is the scaling parameter.
\end{enumerate}
The model based on these assumptions will be referred as the {\bf
M}icro-{\bf C}anonical {\bf E}nsemble with {\bf s}caling {\bf
V}olume {\bf F}luctuations, the {\bf MCE/sVF}.
Calculations presented in this work aim to illustrate the main
idea and therefore they are performed within the simplest possible
model which preserves the essential features: an ideal gas of
massless particles and a mixture of massless neutral with heavy
neutral particles. The Boltzmann approximation has been used.

\vspace{0.3cm} The paper is organized as follows. The description
of the relativistic gas of particles within the MCE  is presented
in Section II.  The scaling volume fluctuations within the MCE
framework are introduced in Section III. Then the basic equations
of the MCE/sVF are given. Next, the results on the multiplicity
distribution, the single particle momentum spectrum and the mean
multiplicity of heavy particles are obtained. The paper ends with
the summary and closing remarks presented in Section IV.

\section {Micro-Canonical Ensemble}
The MCE partition function for the system with $N$ Boltzmann
massless neutral particles reads \cite{Fermi,mce1}:
 \eq{\label{W(N)}
 W_N(E,V)
 = \frac{1}{N!}\left(\frac{gV}{2\pi ^2}\right)^N\int_0^{\infty}\!
 p_1^2dp_1\ldots \int_0^{\infty}\! p_N^2dp_N
       ~\delta(E-\sum_{i=1}^N p_i)
=\frac{1}{E}~\frac{A^N}{(3N-1)!N!}~,
 }
where $E$ and $V$ are the system energy and volume, respectively,
$g$ is the degeneracy factor, and $A \equiv gVE^3/\pi^2$. The MCE
partition function~(\ref{W(N)}) includes exact energy
conservation, but neglects the momentum conservation. The MCE
multiplicity distribution is given by:
\eq{\label{PN-mce}
P_{mce}(N;E,V)~=~\frac{W_N(E,V)}{W(E,V)}~.
}
$W(E,V)$ is the total MCE partition function \cite{mce1}:
\begin{align}\label{WE}
W(E,V) \;\equiv \; \sum_{N=1}^{\infty}W_N(E,V)
     \;=\; \frac{A}{2E}\;\;_0F_3
     \left(;\,\frac{4}{3},\frac{5}{3},2;\,\frac{A}{27}\right)~,
\end{align}
where $_0F_3$ is the generalized hyper-geometric function (see
Appendix). For $A\gg 1$ the mean multiplicity equals to
\cite{mce1}:
\eq{\label{N-mce}
\langle N\rangle_{mce} ~\equiv~
\sum_{N=1}^{\infty}N~P_{mce}(N;E,V)~\cong~(A/27)^{1/4}~,
}
where $P_{mce}(N;E,V)$ was approximated by the normal distribution
\cite{mce1}:
 \eq{\label{PN-mceg}
P_{mce}(N;E,V)~\cong ~(2\pi\, \omega_{mce}\cdot \langle
N\rangle_{mce})^{-1/2}~\exp\left[~-~\frac{\left(~N~-~\langle
N\rangle_{mce}~\right)^2}{2\,\omega_{mce}\cdot\langle
N\rangle_{mce}}~\right] ~,
}
with $\omega_{mce}\equiv(\langle N^2\rangle_{mce}-\langle
N\rangle_{mce}^2)/\langle N\rangle_{mce}=1/4$. Note that in the
grand canonical ensemble (GCE) the multiplicity distribution is
equal to the Poisson one:
 \eq{\label{Pgce}
 P_{gce}(N;T,V) \;=\;
 \frac{\overline{N}^N}{N!}\exp(-\overline{N})\;,
 }
where $\overline{N}$ is the mean multiplicity in the GCE. The
distribution (\ref{Pgce}) approaches the Gaussian for large
$\overline{N}$:
 \eq{
 P_{gce}(N;T,V)
 \;\cong\; (2\pi\, \omega_{gce}\cdot \overline{N})^{-1/2}~
 \exp\left[~-~\frac{\left(~N~-~\overline{N}~\right)^2}
 {2\,\omega_{gce}\cdot\overline{N}}~\right] ~,
 }
with
$\omega_{gce}\equiv(\overline{N^2}-\overline{N}^2)/\overline{N}=1$.
%
%The Eq.~(\ref{N-mce}) is  valid at $\langle N\rangle_{mce} \gg 1$~, and
%Eq.~(\ref{PN-mceg}) requires additionally, $|N-\langle N\rangle_{mce}
%|\ll \langle N\rangle_{mce}$.
%

The numerical calculations presented in this paper will be
performed for $g = 1$ and the energy density which corresponds to
the temperature parameter $T = 160$~MeV. The latter relates the
values of $E$ and $V$ via equation:
\eq{\label{E}
E~=~\frac{3}{\pi^2}~V~T^4~.
}
 The mean multiplicity $\langle N\rangle_{mce}$ in the MCE
(\ref{N-mce}) is then approximately equal to the  GCE value:
\eq{\label{N-ov}
\overline{N} ~=~\frac{1}{\pi^2}~V~T^3~.
}
The approximation $\langle N\rangle_{mce} \cong \overline{N}$ is
valid for $\overline{N}\gg 1$ and reflects the thermodynamic
equivalence of the MCE and the GCE. The scaled variance of the MCE
distribution is $\omega_{mce}=1/4$ \cite{mce1}, and is
approximately independent of $\langle N\rangle_{mce}$ already for
$\langle N\rangle_{mce} > 5$. Thus, despite of thermodynamic
equivalence of the MCE and GCE the value of $\omega_{mce}$ is four
times smaller than the scaled variance of the GCE (Poisson)
distribution, $\omega_{gce} = 1$.

\begin{figure}[ht!]
\epsfig{file=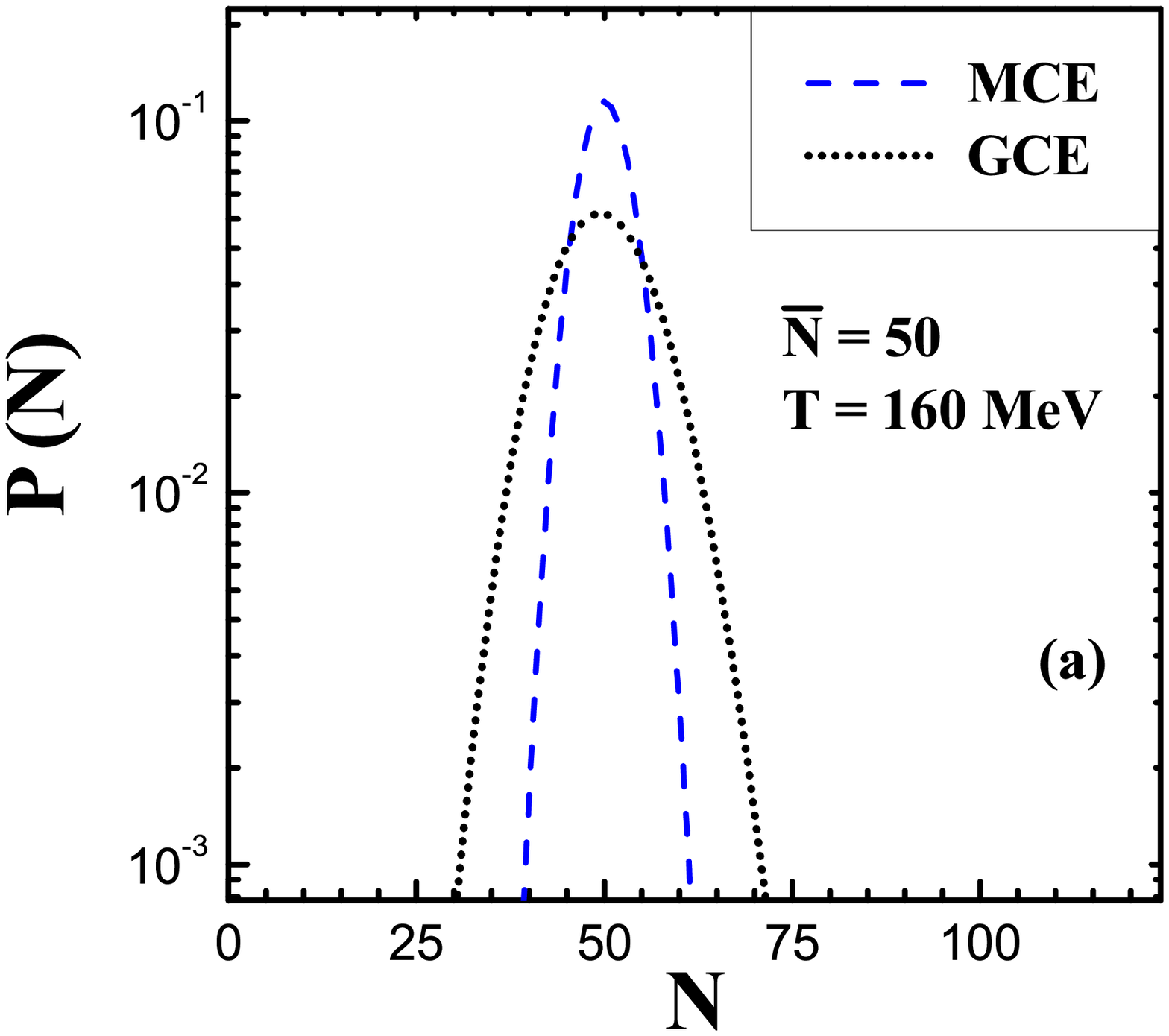,width=0.49\textwidth}\;\;
\epsfig{file=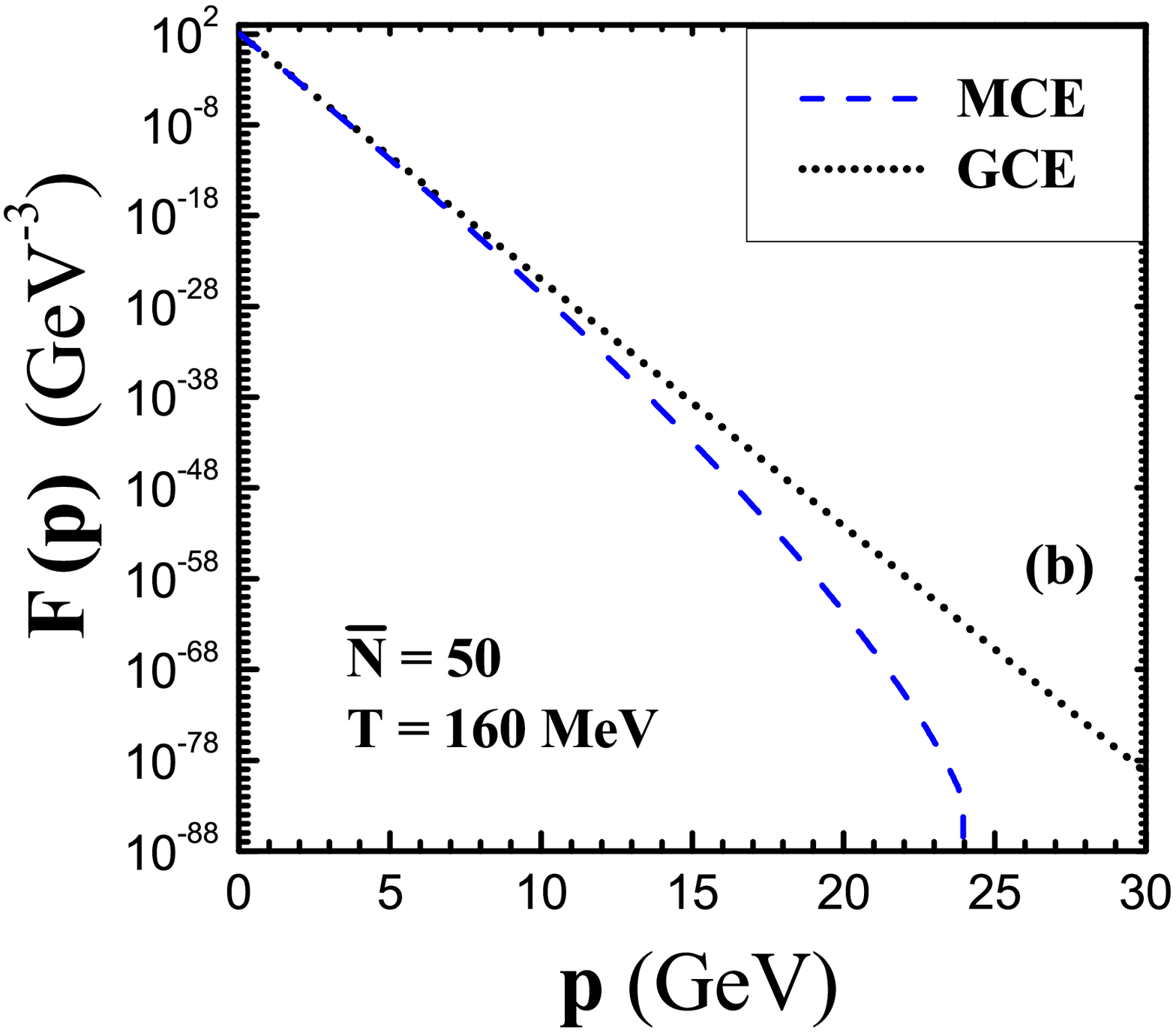,width=0.49\textwidth}
\caption{(Color online) {\bf (a):} The multiplicity distribution
of massless neutral particles in the MCE (\ref{PN-mce}), dashed
line, and the GCE (\ref{Pgce}), dotted line.
% for the same mean multiplicity, $\langle
%N\rangle = 50$,
{\bf (b):} The momentum spectrum of massless neutral particles
calculated within the MCE (\ref{MCE-p}), dashed line, and the GCE
(\ref{Boltz}), dotted line. The system energy is
$E=3\overline{N}T=24$ GeV for both plots.
% for the same mean multiplicity, $\langle
%N\rangle = 50$ and the temperature, $T = 160$~MeV.
 } \label{fig-p-distr}
\end{figure}
%

%-------------------------------------------------------------
%\vspace{0.2cm} \noindent {\bf 8.}
The single particle momentum spectrum in the GCE
reads:
\eq{\label{Boltz}
F_{gce}(p)~\equiv~\frac{1}{\overline{N}}
~\frac{dN}{p^2dp}~=~\frac{V}{2\pi^2~\overline{N}}~\exp\left(-~\frac{p}{T}\right)
~=~\frac{1}{2T^3}~\exp\left(-~\frac{p}{T}\right)~,
}
whereas the corresponding spectrum in the MCE is given by:
\eq{\label{MCE-p}
F_{mce}(p)~&\equiv~\frac{1}{\langle N\rangle_{mce}}~
\frac{dN}{p^2dp}
 ~=~ \frac{V}{2\pi^2\langle N\rangle_{mce}}~
  \sum_{N=2}^{\infty}~ \frac{W_{N-1}(E-p,V)}{W(E,V)}
 ~\equiv~\frac{V}{2\pi^2\langle N\rangle_{mce}}~f(p;E,V)
 \nonumber
 \\
 &=~ \frac{1}{\langle N\rangle_{mce}}~\frac{1}{2E^3}\;
 \sum_{N=2}^{\infty}~\frac{N\,(3N-1)!}{(3N-4)!}
 ~\left(1-\frac{p}{E}\right)^{3N-4}P_{mce}(N;E,V) \; ,
 }
where $f(p;E,V)$ is the MCE analogue of the Boltzmann factor,
$\exp(-p/T)$. From Eq.~(\ref{MCE-p}) follows  $f(p=0;E,V)=1$~.
Both (\ref{Boltz}) and (\ref{MCE-p}) are normalized such that
$\int_0^{\infty}p^2dp~F_{gce}(p)=1$  and
$\int_0^{E}p^2dp~F_{mce}(p)=1$.

Figure~\ref{fig-p-distr}a shows a comparison of the MCE and GCE
results for the multiplicity distribution and
Fig.~\ref{fig-p-distr}b shows the momentum spectrum for
$\overline{N}=50$. The MCE spectrum is close to the Boltzmann
distribution (\ref{Boltz})
%$\exp(-p/T)/(2T^3)$,
at low momenta. This can be shown analytically  using the
asymptotic form of the generalized hyper-geometric function (see
Eq.~(\ref{F03}) in Appendix) at $E\rightarrow \infty$ and $p/E\ll
1$:
\eq{\label{mce-p-as}
f(p;E,V)~=~\frac{W(E-p,V)}{W(E,V)}
 ~=~ \frac{(E-p)^2}{E^2}\,
 \frac{_0F_3\left(;\,\frac{4}{3},\frac{5}{3},2;\,\frac{V(E-p)^3}{27\pi^2}\right)}
      {_0F_3\left(;\,\frac{4}{3},\frac{5}{3},2;\,\frac{VE^3}{27\pi^2}\right)}
 ~\cong~\exp\left(-~\frac{p}{T}\right)~.
}
The MCE spectrum decreases faster than the GCE one at high
momenta. Close to the threshold momentum, $p=E$, where the MCE
spectrum goes to zero, large deviations from (\ref{Boltz}) are
observed. In order to demonstrate this the MCE and GCE momentum
spectra are shown in Fig.~\ref{fig-p-distr}b over 90 orders of
magnitude.

%
% One can also
%easily check that
%
% \eq{\label{F-norm}
% \frac{V}{2\pi^2}\int_0^{E}f(p;E,V)\;p^2dp
% \;=\; \frac{1}{W(E,V)}\sum_{N=2}^{\infty}W_N(E,V)~ N
% \;\cong \; \langle N\rangle_{mce}\;.
% }
%
%The normalization condition (\ref{F-norm}) becomes valid at
%$\langle N \rangle_{mce} \gg 1$~.
%{\bf I do not understand this, it seems to be in contradiction with
%Eq.~16, where there are no approximate relations!!!}

%------------------------------------------------------
%\vspace{0.2cm} \noindent {\bf 9.}
\section{MCE with scaling Volume Fluctuations}
Let us consider a set of micro-canonical ensembles with the same
energy $E$ but different volumes $V$.
%All micro-states which belong to
%this set constitute a new ensemble which is introduced in
%Ref.~\cite{alpha} and called the $\alpha$-ensemble.
The probability density which describes the volume fluctuations is
denoted by $P_{\alpha}(V)$. The distribution of any quantity $X$
can then be calculated as:
\eq{\label{a-ensemble}
P_{\alpha}(X;E)~=~\int_0^{\infty}
dV~P_{\alpha}(V)~P_{mce}(X;E,V)~,
}
where $P_{mce}(X;E,V)$ is the distribution of the quantity $X$ in
the MCE with fixed $E$ and $V$. Further more it is assumed that
the distribution $P_{\alpha}(V)$ has the scaling form:
\eq{\label{scale} P_{\alpha}(V)=\overline{V}^{~-1}~
\phi_{\alpha}(V/\overline{V})~,
 }
with the volume parameter $\overline{V}$ being the scale parameter
of $P_{\alpha}(V)$. From~Eq.~(\ref{scale}) follows that the scale
parameter is proportional to the average volume and the
proportionality factor follows from the normalization conditions
of the scaling function (see below). Equations (\ref{a-ensemble})
and (\ref{scale}) are the basic equations of the MCE/sVF.
% defined by the assumptions (AI)-(AIV).

Let us introduce an auxiliary variable $y$ defined as
$y~\equiv~\left(V/\overline{V}\right)^{1/4}$. Using
Eqs.~(\ref{N-mce}) and (\ref{E}, \ref{N-ov}) the mean multiplicity
of massless particles in the MCE (\ref{N-mce}) can be written as:
 \eq{\label{N-av}
 \langle N \rangle_{mce}~=~
\overline{N}
 ~y~,
 }
where
\eq{\label{V-av}
\overline{N}~=~\left(\frac{E^3\overline{V}}{27\pi^2}\right)^{1/4}~.
}
The average temperature $\overline{T}$ is related (see
Eq.~(\ref{E})) to the energy and volume parameter $\overline{V}$
as:
 \eq{\label{E1}
E~=~\frac{3}{\pi^2}~\overline{V}~\overline{T}^4~.
}
Within this paper $\overline{T}=160$~MeV is assumed. The change of
the volume at $E=const$ leads to the corresponding change of the
system temperature:
 \eq{\label{T}
 T\;=\;\overline{T}\cdot(\overline{V}/V)^{1/4}=\overline{T}/y.
 }
The temperature fluctuations resulting from the volume
fluctuations will be discussed below.

 Using the variable $y$, the volume integral in
Eq.~(\ref{a-ensemble}) can be conveniently rewritten as:
\eq{\label{a-ensemble1}
P_{\alpha}(X;E)=\left(\overline{V}\right)^{-1}\int_0^{\infty}
dV\phi_{\alpha}\left(V/\overline{V}\right)P_{mce}(X;E,V)=\int_0^{\infty}
dy~\psi_{\alpha}(y)P_{mce}(X;E,y^4\overline{V})~,
}
where $\psi_{\alpha}(y)\equiv 4y^3 \phi_{\alpha}(y^4)$. Choosing
$\psi_{\alpha}(y)=\delta(y-1)$ one recovers the MCE with
$V=\overline{V}$ and $\langle N\rangle_{mce} =\overline{N}$.

The
scaling function $\psi_{\alpha}(y)$ will be required to satisfy
two normalization conditions:
 \eq{\label{int-psi}
 \int_0^{\infty}dy~\psi_{\alpha}(y)\;=\;1\;,~~~~
 %\\
 \int_0^{\infty}dy~y\,\psi_{\alpha}(y) \;=\;1\;.
% \label{int-z-psi}
 }
The first condition guarantees the proper normalization of the
volume probability density function,
$\int_0^{\infty}dVP_{\alpha}(V)=1$. The second condition is
selected in order to keep the mean multiplicity in the MCE/sVF
equal to the  MCE  mean multiplicity.

%---------------------------------------------------------------
\subsection{KNO-scaling}
The multiplicity distribution in the MCE/sVF is:
 \eq{\label{P-N-alpha}
 P_{\alpha}(N; \overline{N})~=~\int_0^{\infty}dV~P_{\alpha}(V)~P_{mce}(N;E,V)~=~
 \int_0^{\infty}dy~\psi_{\alpha}(y)\,P_{mce}(N;E,y^4\overline{V})\;.
 }
At $\overline{N}\gg 1$, the particle number distribution
(\ref{P-N-alpha}) can be approximated as:
 \eq{\label{PNz}
 P_{\alpha}(N;\overline{ N})
 \;\cong\; \langle N\rangle_{\alpha}^{-1}~\psi_{\alpha}(z)~,
 }
where $z\equiv N/\langle N \rangle_{\alpha}$\;, and the mean
multiplicity $\langle N\rangle_{\alpha}$ is given by:
\eq{\label{N-a}
\langle N\rangle_{\alpha}~=~\sum_{N=1}^{\infty}~ N~P_{\alpha}(N;
\overline{N})~\cong~\overline{N}~\int_0^{\infty}dy~y~\psi_{\alpha}(y)~
=~\overline{N} ~.
}
The approximate equality of $\langle N\rangle_{\alpha}$ and
$\overline{N}$ is satisfied for $\overline{N}\gg 1$ due to the
second normalization condition  (\ref{int-psi}).
The KNO scaling of the multiplicity distribution
$P_{\alpha}(N;\overline{ N})$ follows from the assumption of the
scaling of the volume fluctuations (\ref{scale}).
%The multiplicity scaling function equals to the function
%$\psi_{\alpha}$.

For convenience,  a simple analytical form of the scaling function,
$\psi_{\alpha}$ will be used:
\eq{\label{psi-a}
 \psi_{\alpha }(y) \;=\; \frac{k^k}{\Gamma(k)}~y^{k-1}~\exp(-k\,y)\;,
 }
where $\Gamma(k)$ is the Euler gamma function.  It was found
\cite{stas} that the function (\ref{psi-a}) with $k=4$
approximately describes the experimental data on KNO scaling in
p+p interactions. All numerical calculations obtained within the
MCE/sVF and presented in this paper will be done using the
function (\ref{psi-a}) with $k=4$. Note, that the function
(\ref{psi-a}) satisfies both normalization conditions
(\ref{int-psi}) for any $k>0$.
In order to check the sensitivity of the results derived within
the MCE/sVF to the shape of the KNO scaling function, the
calculations presented below were repeated using the scaling
function resulting from the fit to the experimental
data~\cite{knog1}:
\eq{\label{psi-b}
 \psi_{\alpha}(y) \;=\; a~ y^c~\exp(-b\,y^2) \;,
 }
where the values of the parameters are the following: $a = 1.19$,
$b = 0.62$ and $c = 0.66$.

\begin{figure}[ht!]
\epsfig{file=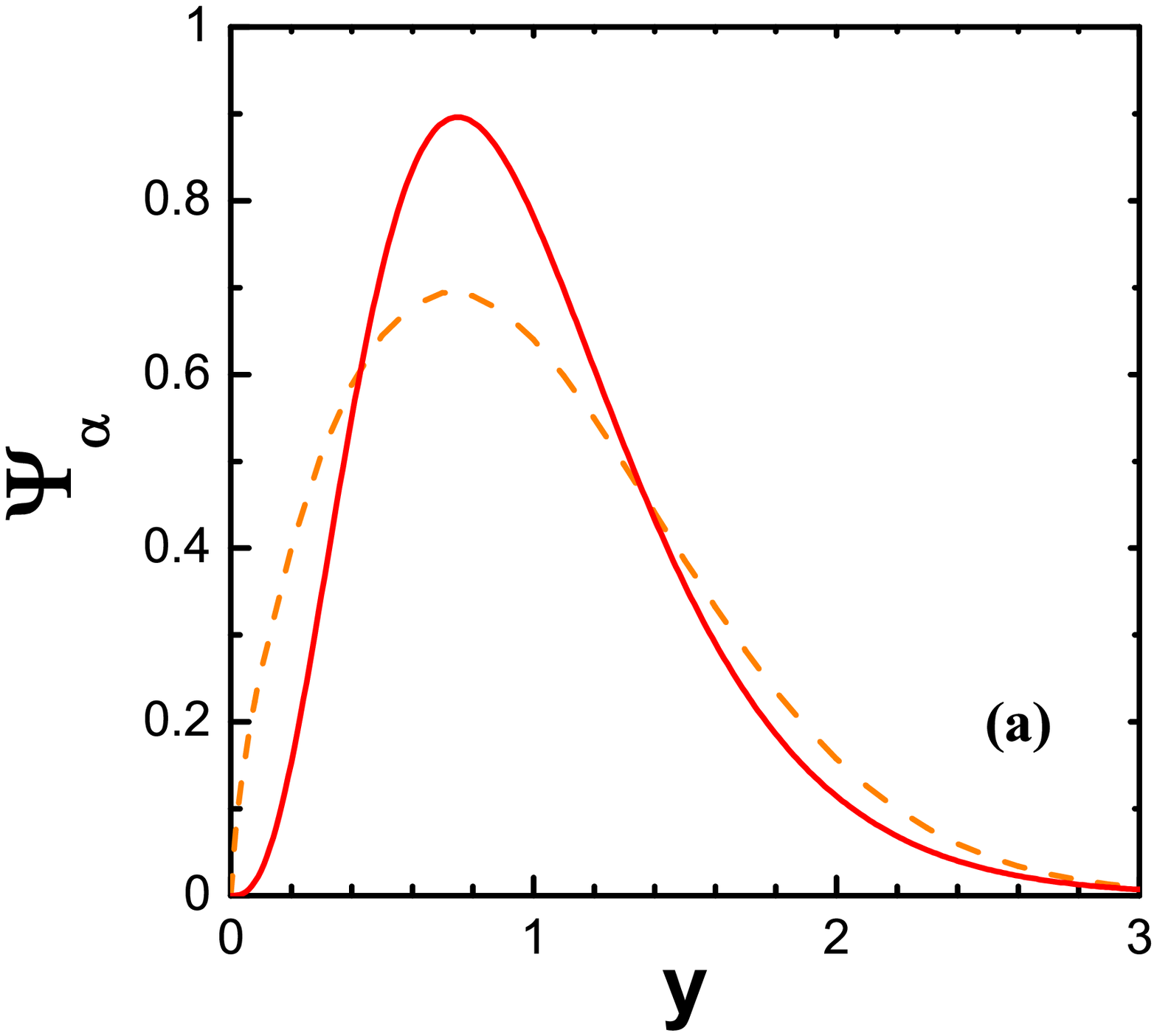,width=0.49\textwidth}\;\;
\epsfig{file=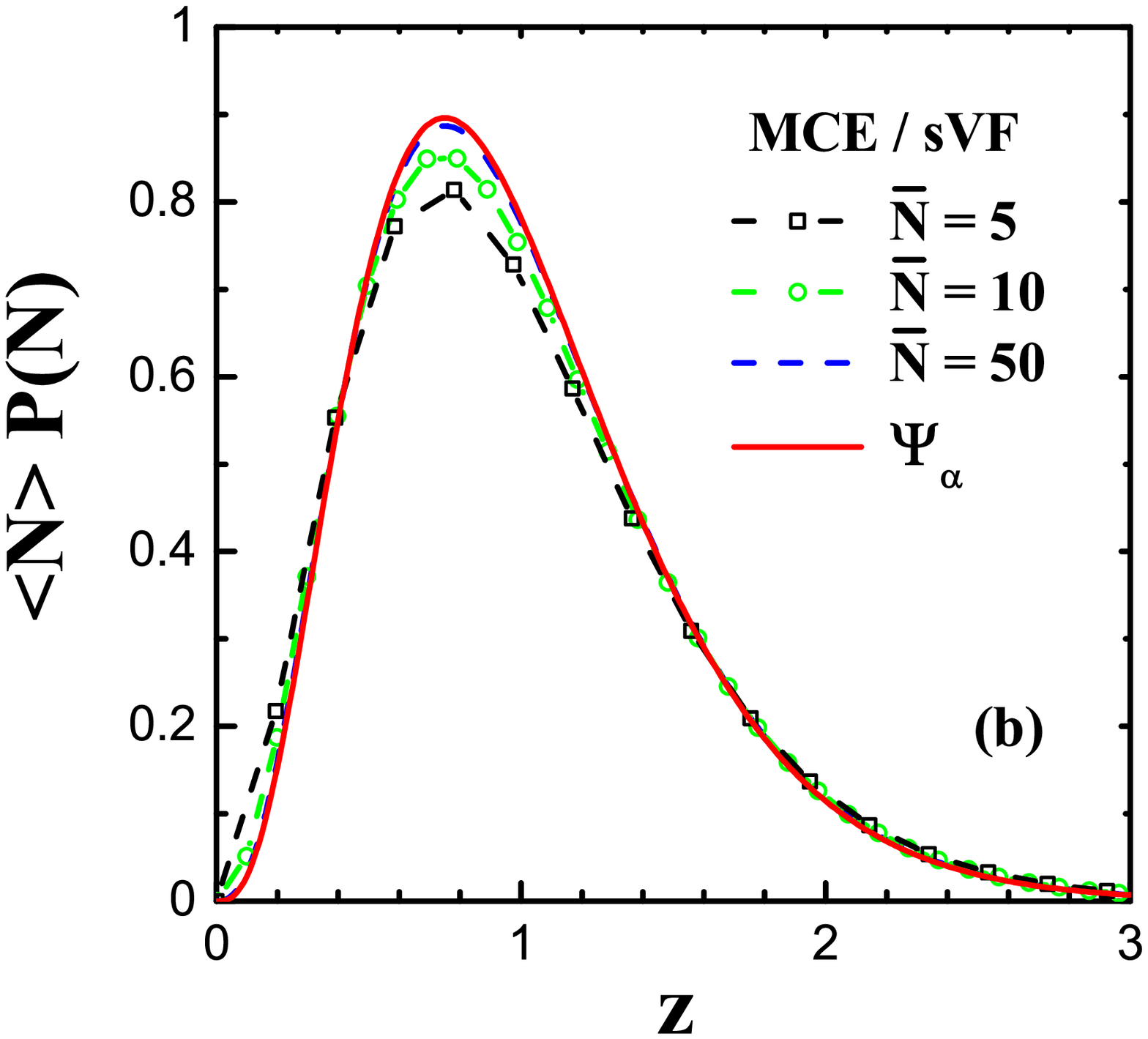,width=0.49\textwidth}
 \caption{(Color online) {\bf (a):}
The scaling functions determining the volume fluctuations
discussed in this paper. Solid line indicates the function
(\ref{psi-a}) with $k=4$ for which all presented results are
calculated, whereas the dashed line shows the function
(\ref{psi-b}) used for the sensitivity tests. {\bf (b):} The
multiplicity distributions calculated within the MCE/sVF for three
values of mean multiplicity. The calculations are performed for
the function (\ref{psi-a}) with $k=4$. The distributions are
presented in the KNO variables, $\langle N\rangle\cdot P(N)$ and
$z=N/\langle N\rangle$.
 } \label{fig-psi}
\end{figure}

In Fig.~\ref{fig-psi}a both KNO scaling functions (\ref{psi-a})
and (\ref{psi-b}) are plotted for a comparison. The qualitative
behavior is similar, however there are substantial quantitative
differences. Nevertheless,  the main results presented below are
the same for both scaling functions.

The multiplicity distributions calculated within the MCE/sVF for
the three mean multiplicities, $\overline{N} =$ 5, 10 and 50, are
shown in Fig.~\ref{fig-psi}b in the KNO variables. The scaling
function (\ref{psi-a}) is also shown for a comparison. These
numerical results indicate that even for relatively low mean
multiplicities the KNO scaling is approximately obeyed by the
MCE/sVF distributions.

%%% tutaj
The KNO scaling of the multiplicity distribution implies that the
scaled variance of the distribution increases in proportion to the
mean multiplicity.
%This holds for the MCE/sVF
%model.
For $\overline{N}\gg 1$ one gets:
\eq{\label{omega-alpha}
\omega_{\alpha}~=~\frac{\langle N^2\rangle_{\alpha}~-~\langle
N\rangle_{\alpha}^2}{\langle
N\rangle_{\alpha}}~\cong~\kappa~\langle N \rangle_{\alpha}~,
}
where $\kappa =const >0$. Equation~(\ref{omega-alpha}) results
from an approximation of $\langle N^2\rangle_{\alpha}$ for
$\overline{N}\gg 1$, namely:
\eq{
\langle N^2\rangle_{\alpha}~=~\sum_{N=1}^{\infty} N^2
~P_{\alpha}(N;\overline{N})~\cong~
\overline{N}^2~\int_0^{\infty}dy~y^2~\psi_{\alpha}(y)~
\cong~(1+\kappa)~\langle N \rangle_{\alpha}^2~.
\label{N2}
 }
The positive value of $\kappa$ in Eq.~(\ref{N2}) follows from
the normalization conditions (\ref{int-psi}):
\eq{\label{kappa}
\kappa~=~\int_0^{\infty}dy~y^2~
\psi_{\alpha}(y)~-~1~=~\int_0^{\infty}dy~(y-1)^2~\psi_{\alpha}(y)~>~0~,
}
For the function $\psi_{\alpha}(y)$ defined by Eq.~(\ref{psi-a})
one finds, $\kappa = k^{-1}$.

\begin{figure}[ht!]
\epsfig{file=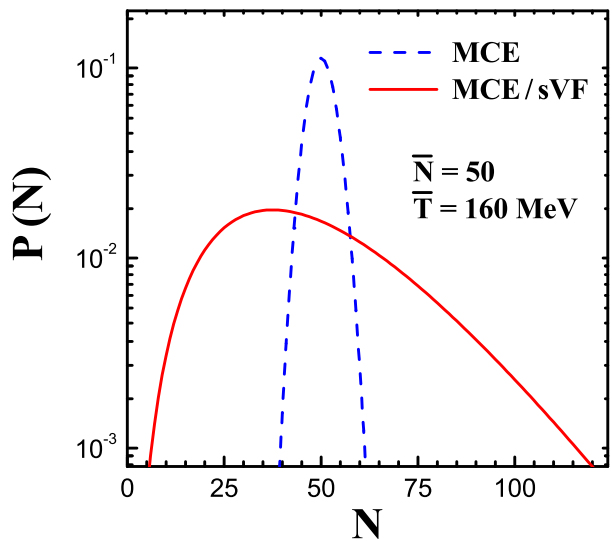,width=0.49\textwidth}
 \caption{(Color online) A comparison of the multiplicity distributions
 of massless neutral particles calculated with the MCE/sVF
 (solid line) and the MCE (dashed line). The system energy is
$E=3\overline{N}\,\overline{T}=24$ GeV.
% for the same mean multiplicity, $\langle N \rangle = 50$.
 } \label{fig-mce-psi}
\end{figure}

In Fig.~\ref{fig-mce-psi} the multiplicity distributions obtained
within the MCE and the MCE/sVF for $\overline{N}=50$ are compared.
The scaled variance of the MCE/sVF distribution for
$\overline{N}=50$ is about 12.5, whereas the scaled variance of
the MCE distribution is 1/4. This large difference in the width of
the MCE/sVF and the MCE distributions is clearly seen in the
figure.

The volume averaging procedure results in the mean value given by:
\eq{\label{V-alpha}
\langle V\rangle_{\alpha}~=~\int_0^{\infty}dV~V~P_{\alpha}(V)~=~
\overline{V}~\int_0^{\infty}dy~y^4~\psi_{\alpha}(y)~\equiv~a~\overline{V}~.
}
For $\psi_{\alpha}$ (\ref{psi-a})  one gets:
\eq{\label{a}
a~=~\frac{(k+1)(k+2)(k+3)}{k^3}~,
}
 which gives $a\cong 3.28$ for $k=4$. Thus, in
general, the mean volume in the MCE/sVF is not equal to the
corresponding MCE volume.
%Consequently, the MCE/sVF and the MCE
%results are different even for the mean quantities in the large
%volume limit.
%%% tutaj
%---------------------------------------------------------------
\subsection{Power Law in Momentum Spectrum}
The volume fluctuations in the MCE/sVF significantly increase the
width of the multiplicity distribution. They are also expected to
modify the single particle momentum spectrum. This is because for
a fixed system energy, the volume of the system determines the
energy density, and consequently, the effective temperature of
particles.

The single particle momentum spectrum within the MCE/sVF
can be directly calculated from
Eq.~(\ref{a-ensemble}) and it reads:
 \eq{\label{a-spectrum}
F_{\alpha}(p)~&\equiv~\frac{1}{\langle N\rangle_{\alpha}}~{\Big
\langle} \frac{dN}{p^2dp}{ \Big \rangle}_{\alpha}~=~
\frac{1}{\langle
N\rangle_{\alpha}}~\int_0^{\infty}dV~P_{\alpha}(V)~
\frac{V}{2\pi^2}~f(p;E,V)
\nonumber \\
% &
 &=\; \frac{1}{\langle N\rangle_{\alpha}}~\frac{1}{2E^3}
 \sum_{N=2}^{\infty} \frac{N~(3N-1)!}
 {(3N-4)!}~\left(1~-~\frac{p}{E}\right)^{3N-4}~P_{\alpha}(N;\overline{N})~.
 }
The formal structure of the expression (\ref{a-spectrum})  is
similar to the structure of the corresponding expression derived
within the MCE (\ref{MCE-p}). The only, but the crucial,
difference is that the narrow MCE multiplicity distribution used
for averaging the particle spectrum in Eq.~(\ref{MCE-p}) is
replaced by the broad MCE/sVF multiplicity distribution in
Eq.~(\ref{a-spectrum}).  The spectrum $F_{\alpha}(p)$ fulfills the
normalization condition, $\int_0^{E}p^2dp~F_{\alpha}(p)=1$. From
Eq.~(\ref{a-spectrum}) one finds $F_{\alpha}(p=0)=a\cdot
F_{gce}(p=0)$. Equation~(\ref{a}) gives $a \cong 3.28$ for the
$\psi_{\alpha}$ function (\ref{psi-a}) with $k=4$. Thus, in the
MCE/sVF there is an enhancement of the momentum spectrum at
$p\rightarrow 0$ compared to the GCE and MCE results.

\begin{figure}[ht!]
\epsfig{file=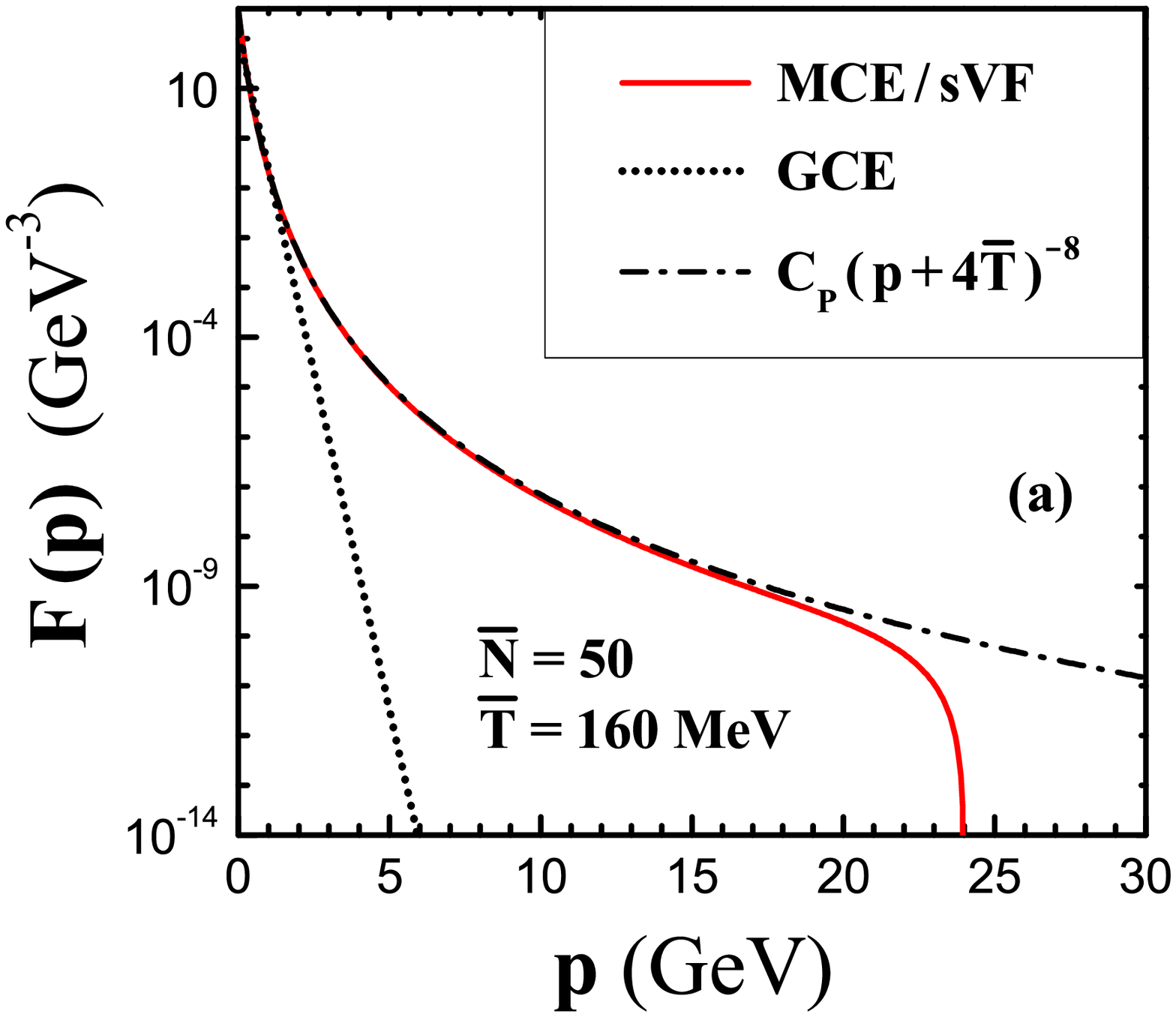,width=0.49\textwidth}\;\;
\epsfig{file=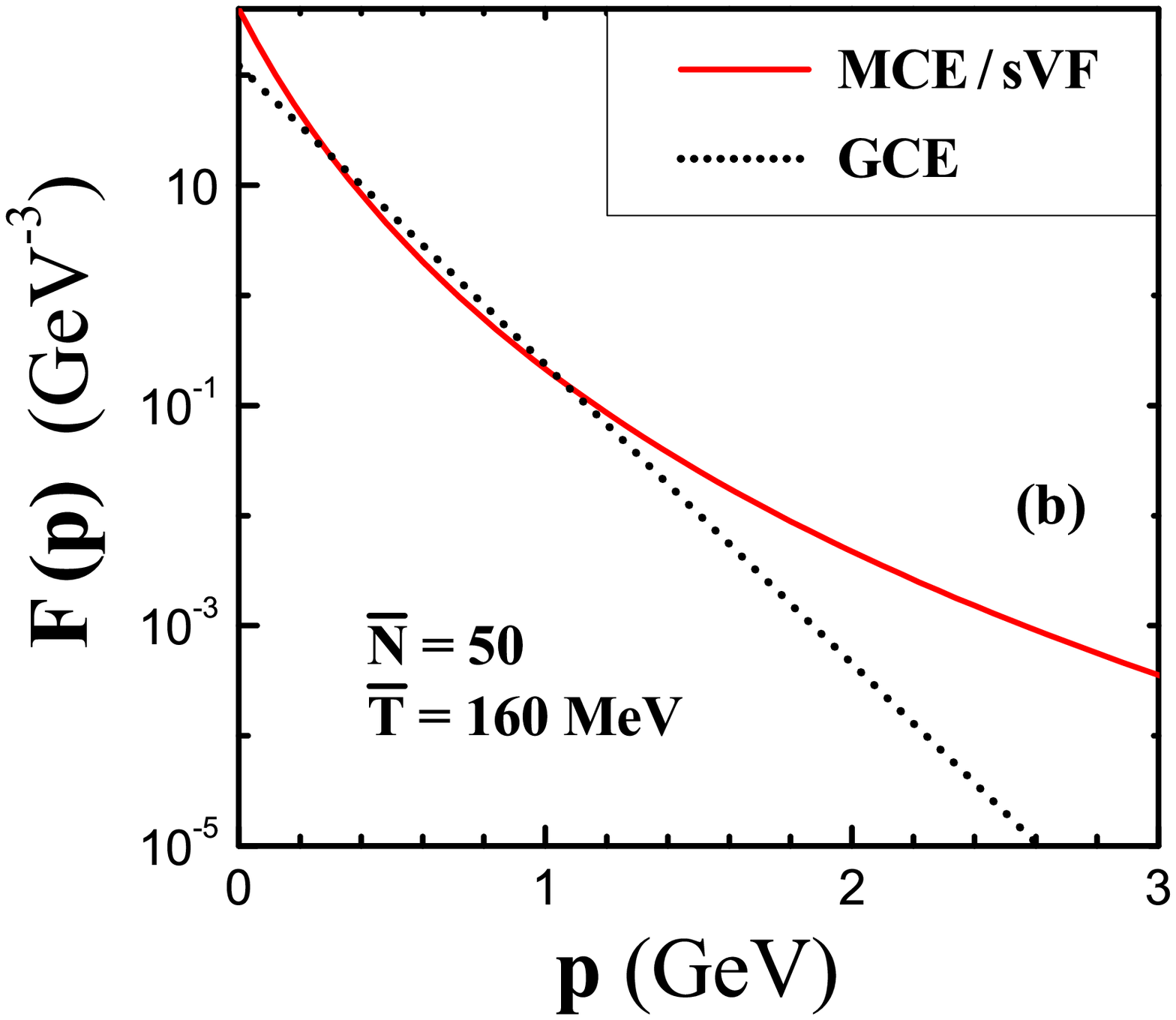,width=0.49\textwidth}
\caption{(Color online) {\bf (a):} The momentum spectrum of
massless neutral particles calculated within the MCE/sVF
(\ref{a-spectrum}), solid line, and the GCE (\ref{Boltz}), dotted
line. The approximation~(\ref{ana1}) of the MCE/sVF spectrum is
shown by the dashed-dotted line. {\bf (b):} The same as in the
left panel but only for the low momentum region. The
approximation~(\ref{ana1}) is not shown as it overlaps with the
MCE/sVF line. The system energy is
$E=3\overline{N}\,\overline{T}=24$ GeV for both plots.}
\label{fig-momentum}
\end{figure}

The single particle momentum spectrum calculated with the MCE/sVF
for the volume scaling function (\ref{psi-a}) is shown in
Fig.~\ref{fig-momentum}a. A striking new feature of this spectrum
is the presence of a long power law tail. In the momentum range
from several GeV to about 20 GeV the spectrum can be approximated
by:
\eq{\label{Kp}
F_{\alpha}(p)~\cong~C_p~p^{-K_p}~,
}
with $C_p$ and $K_p$ being the normalization and power parameters,
respectively. For momenta smaller than 3~GeV the spectrum starts
to deviate significantly from the power law parametrization and
its local inverse slope parameter is close to the temperature of
the corresponding GCE, $\overline{T} = 160$~MeV. This is shown in
Fig.~\ref{fig-momentum}b, where the momentum spectrum for $ p \le
3$~GeV is presented. A rapid decrease of the spectrum starts at $p
\ge 20$~GeV, when the threshold value $p = 24$~GeV is approached.
The above features of the MCE/sVF momentum spectrum resemble
features of the transverse momentum spectrum of hadrons produced
in high energy p+p interactions~\cite{pp_spectra}.

\begin{figure}[ht!]
\epsfig{file=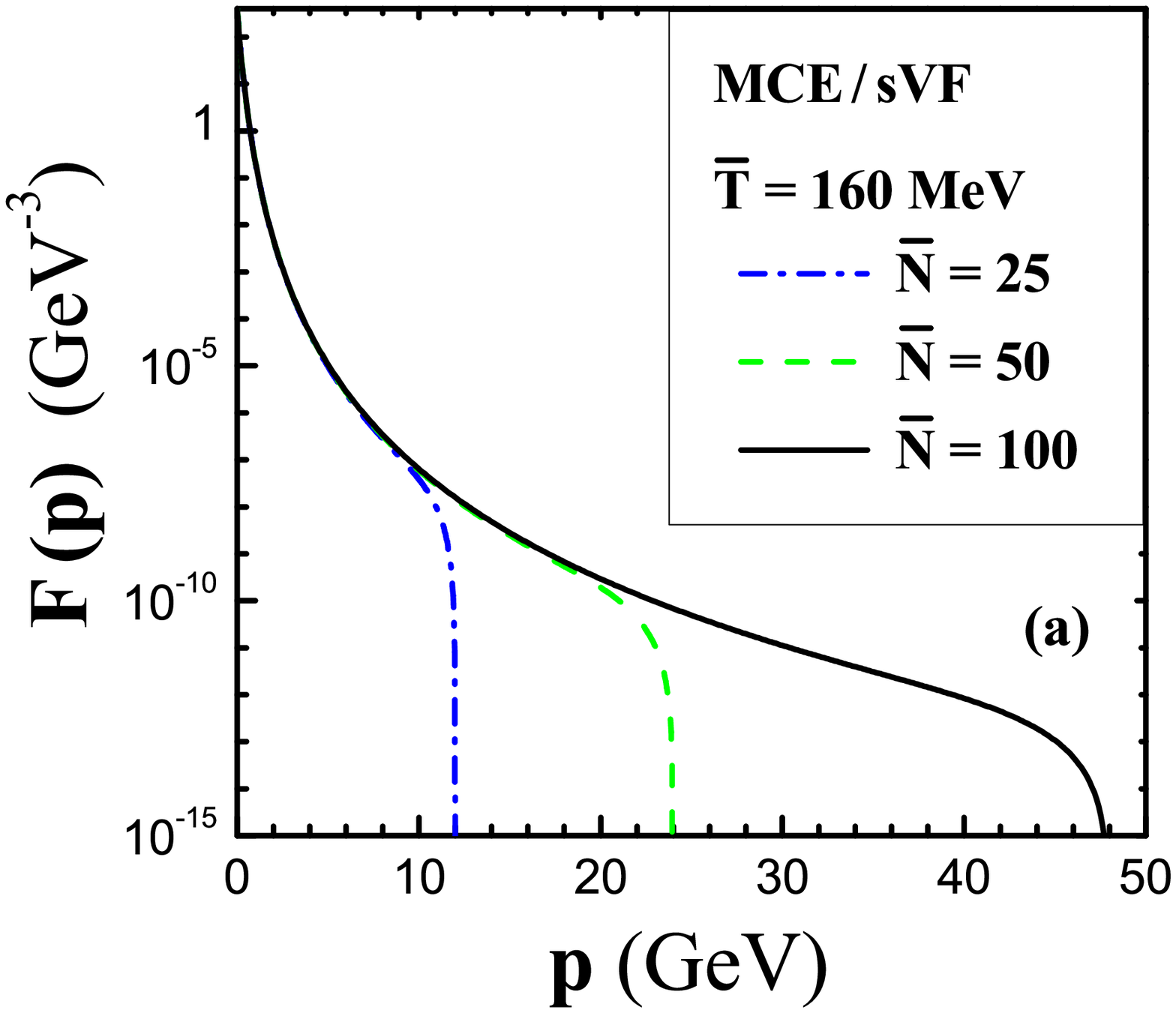,width=0.49\textwidth}\;\;
\epsfig{file=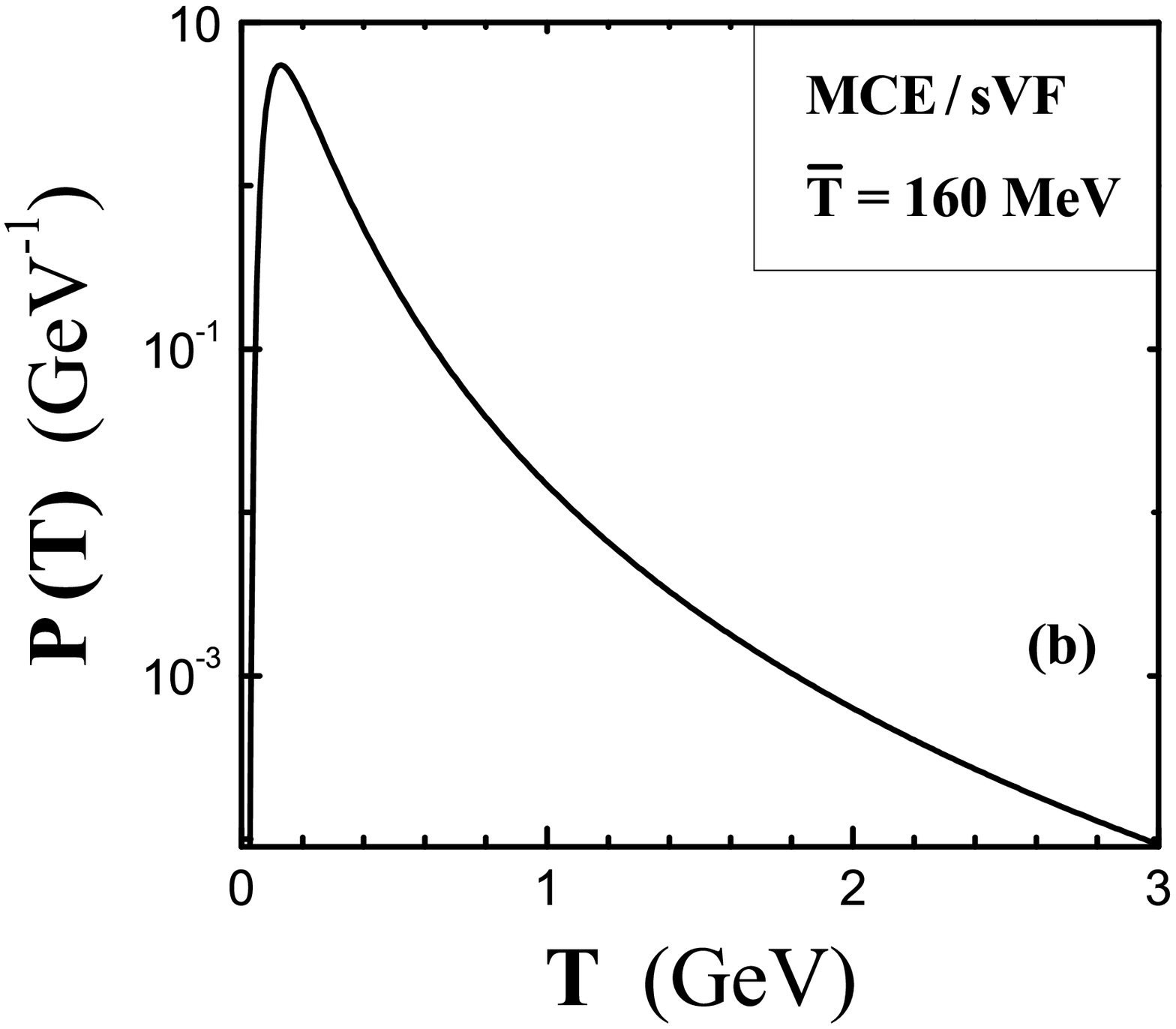,width=0.49\textwidth}
\caption{(Color online) {\bf (a):} The momentum spectrum of
massless neutral particles calculated within the MCE/sVF
(\ref{a-spectrum}) for three values of the mean multiplicity equal
to $\overline{N}=25$, dashed-dotted line, $\overline{N}=50$,
dashed line, and $\overline{N}=100$, solid line. The system energy
is $E=3\overline{N}\,\overline{T}$. {\bf (b):} The temperature
probability distribution (\ref{PT}) with $k=4$ in the MCE/sVF.}
\label{fig-p-N}
\end{figure}

The power law dependence (\ref{Kp}) of the momentum spectrum at
high momenta can be derived analytically, namely:
 \eq{\label{ana1}
  F_{\alpha}(p)
 ~&=~\int_0^{\infty}F_{mce}(p)\,\psi_{\alpha}(y)\,dy~
 \cong~
 %\frac{1}{2\pi^2\,\overline{N}}~\int_0^{\infty}\;dV~V~P_{\alpha}(V)
 %      \exp\left(-~\frac{p}{T}\right)
        %\; \int_0^{\infty}F_{gce}(p)\,\psi_{\alpha}(y)\,dy\\
 % =
 %\nonumber
 \frac{\overline{V}}{2\pi^2\overline{N}}\int_0^{\infty}dy~\psi_{\alpha}(y)
       y^4 \exp\left(-~\frac{p}{\overline{T}}\,y\right)~\nonumber \\
& =~
\frac{k^{k}~\Gamma(k+4)}{2~\Gamma(k)}~\overline{T}^{k+1}~(p+\overline{T}k)^{-k-4}
%        \frac{\Gamma(k+4)}{\Gamma(k)} \left(\sqrt{m^2+p^2}\;+\;k\,T\right)^{-k-4}
%  \;=\; \frac{C_{P}}{\left(\sqrt{m^2+p^2}\;+\;k\,T\right)^{k+4}}
%\nonumber
%&        \\
~ \cong ~11.27~\rm{GeV}^5~ (p+4\overline{T})^{-8}~,
 %\nonumber
%
 }
where $\overline{T}=160$~MeV and $k = 4$ is set in the last
expression. Note that according to Eq.~(\ref{T}) the temperature,
$T=\overline{T}/y$, fluctuates\footnote{In Refs.~\cite{wilk} it
was shown for the first time that the properly chosen temperature
fluctuations may lead to the Tsallis distribution \cite{tsallis}
with power law spectrum at high (transverse) momenta, see also
Ref.~\cite{Wilk:2006vp}.} in the MCE/sVF due to the volume
fluctuations.
% and Eq.~(\ref{ana1}) can
%be interpreted as an averaging over the $T$ fluctuations
The volume fluctuations are responsible for the appearance of the
power law tail. The temperature probability distribution can be
easily derived:
\eq{\label{PT}
P(T)~=~ \frac{\overline{T}}{T^2}\; \psi_{\alpha}(\overline{T}/T)
 \;=\; \frac{1}{\overline{T}}~\frac{k^k}{\Gamma(k)}~
       \left(\frac{\overline{T}}{T}\right)^{k+1}
            \exp\left(-\,k\,\frac{\overline{T}}{T}\right)~.
}
The result is shown in Fig.~\ref{fig-p-N}b.

The analytical
approximation in Eq.~(\ref{ana1}) is valid for
%the particle energies much
%smaller than the system energy,
$p \ll E$. It describes the particle spectra at  $p\ll E$
including the low momentum region (see Fig.~\ref{fig-momentum},
(a)). For $p \gg \overline{T}$ one gets %the parameters of
the power law parameters (\ref{Kp}):  $C_p\cong 11.27$~GeV$^{-5}$
and $K_p=8$. The power law dependence of the particle energy
spectrum appears when there is a distinct region of the particle
energies in which $\overline{T} \ll p \ll E$.

Equation~(\ref{ana1}) leads to an exponential spectrum for $p\ll
\overline{T}$ :
\eq{\label{low-p}
 F_{\alpha}(p)~\cong~
 \frac{a}{2\overline{T}^3}~\exp\left(-~\frac{p}{T_0}\right)~,
 }
with the inverse slope parameter $T_0=\overline{T}/2$ and the
normalization constant $a \cong 3.28$. For arbitrary $k>0$, the
parameter $a$ is given by Eq.~(\ref{a}), and
$T_0=\overline{T}k/(k+4)$.
Within the used approximation the power parameter, $K_p$, depends
only on the form of the volume scaling function, whereas the
normalization constant $C_{p}$ in Eq.~(\ref{Kp}) is sensitive, in
addition, to the temperature $\overline{T}$.

Note that the scaling volume fluctuations at a fixed system energy
lead alone to the KNO scaling and the power law behavior of the
spectra. Systems with small volume, $V\ll \overline{V}$, have
predominately a small particle multiplicity, $N\ll \overline{N}$.
Due to the energy conservation these systems have large
temperatures, $T\gg \overline{T}$, which results in the power law
behavior of the momentum spectra and the mean heavy particle
multiplicity (see the next Section). The energy fluctuations at
fixed volume do not lead to the power law dependence. This is
because, systems with a small energy, $E \ll \overline{E}$, have
predominately a small particle multiplicity, $N\ll \overline{N}$,
and consequently the temperature is  approximately independent of
the system energy, volume or multiplicity. Thus, no power law
behavior appears in this case.

On the other hand, the energy fluctuations can be added into the
present consideration. They do not change the principal behavior
of particle spectra. In order to illustrate the independence of
the power law tail of the system energy the MCE/sVF spectra
calculated for three different values of $E$
% which correspond to $\overline{N} = 25,\; 50$, and 100,
and compared in Fig.~\ref{fig-p-N}a. In the overlapping power law
region (left to the cuts from the energy conservation) the spectra
are similar, i.e. they do not depend on the average multiplicity.
This implies that averaging over collisions with different
energies deposited for particle production would not change the
particle spectra for momenta below the region affected by the
steep threshold decrease. In particular, it follows from
Fig.~\ref{fig-p-N}a  that the energy fluctuations in the region
$E=10\div 50$~GeV do not change the momentum spectra at
$p<10$~GeV. The same arguments are applied to the mean
multiplicity of heavy particles discussed in the next Section and
shown in Fig.~\ref{fig-nM}b

 Finally, it is important to note, that the momentum
spectrum for particles with non-zero mass, $m$, obeys a power law
dependence in the particle energy. This is obvious from
Eq.~(\ref{ana1}), because the particle momentum, $p$, in the
exponent can be replaced by the particle energy, $\sqrt{p^2 +
m^2}$, resulting in:
 \eq{\label{ana1a}
  F_{\alpha}(p)
 &\;\cong~
  \frac{\overline{V}}{2\pi^2\overline{N}}\int_0^{\infty}dy~\psi_{\alpha}(y)
       y^4 \exp\left(-~\frac{\sqrt{p^2 + m^2}}{\overline{T}}\,y\right)~\nonumber \\
 &\;=~
\frac{k^{k}~\Gamma(k+4)}{2~\Gamma(k)}~\overline{T}^{k+1}~(\sqrt{p^2 + m^2}+
\overline{T}k)^{-k-4}~.
%        \frac{\Gamma(k+4)}{\Gamma(k)} \left(\sqrt{m^2+p^2}\;+\;k\,T\right)^{-k-4}
%  \;=\; \frac{C_{P}}{\left(\sqrt{m^2+p^2}\;+\;k\,T\right)^{k+4}}
%\nonumber
%&        \\
%~ \cong ~11.27~\rm{GeV}^5~ (\sqrt{p^2 + m^2}+4\overline{T})^{-8}~,
 %\nonumber
%
 }
Consequently, the dependence of the momentum spectrum on the
particle momentum and mass reduces to the dependence on the
particle energy. This energy scaling predicted by the MCE/sVF
seems to coincide with the scaling in the transverse mass of the
quarkonia spectra measured in p+$\overline {\rm{p}}$ interactions
\cite{powerlaw}.

%---------------------------------------------------------------
%\vspace{0.2cm} \noindent {\bf 11.}
\subsection{ Power Law in Mass Dependence }
The volume fluctuations together with the specific scaling
function (\ref{psi-a}) lead to the power law dependence of the
momentum spectrum at high momenta. It may be expected that these
volume fluctuations will also cause a similar dependence of mean
multiplicity of heavy particles on the particle mass. This subject
is discussed below in the approximation of heavy and thus
non-relativistic particles.

\begin{figure}[ht!]
\epsfig{file=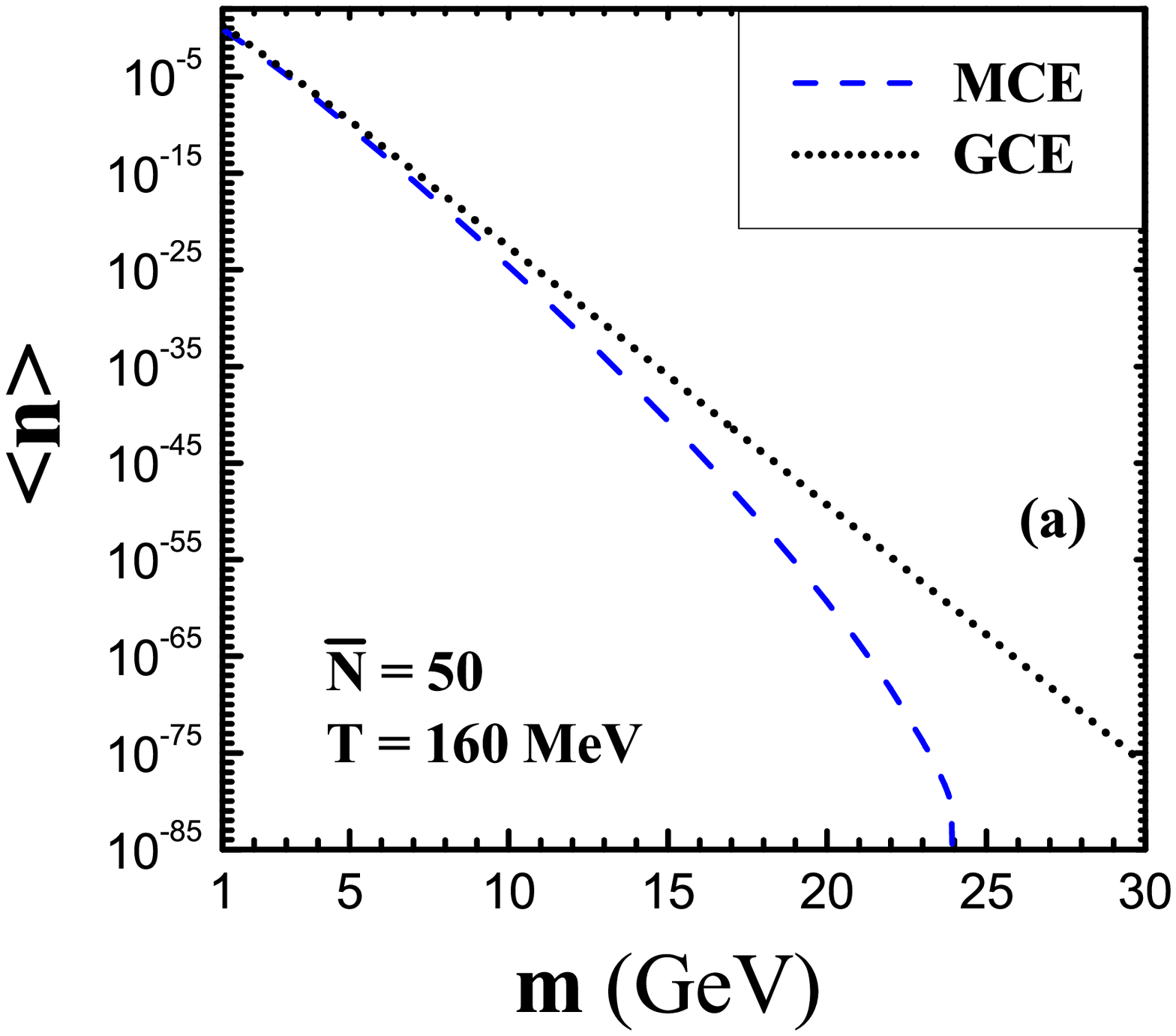,width=0.49\textwidth}\;\;
\epsfig{file=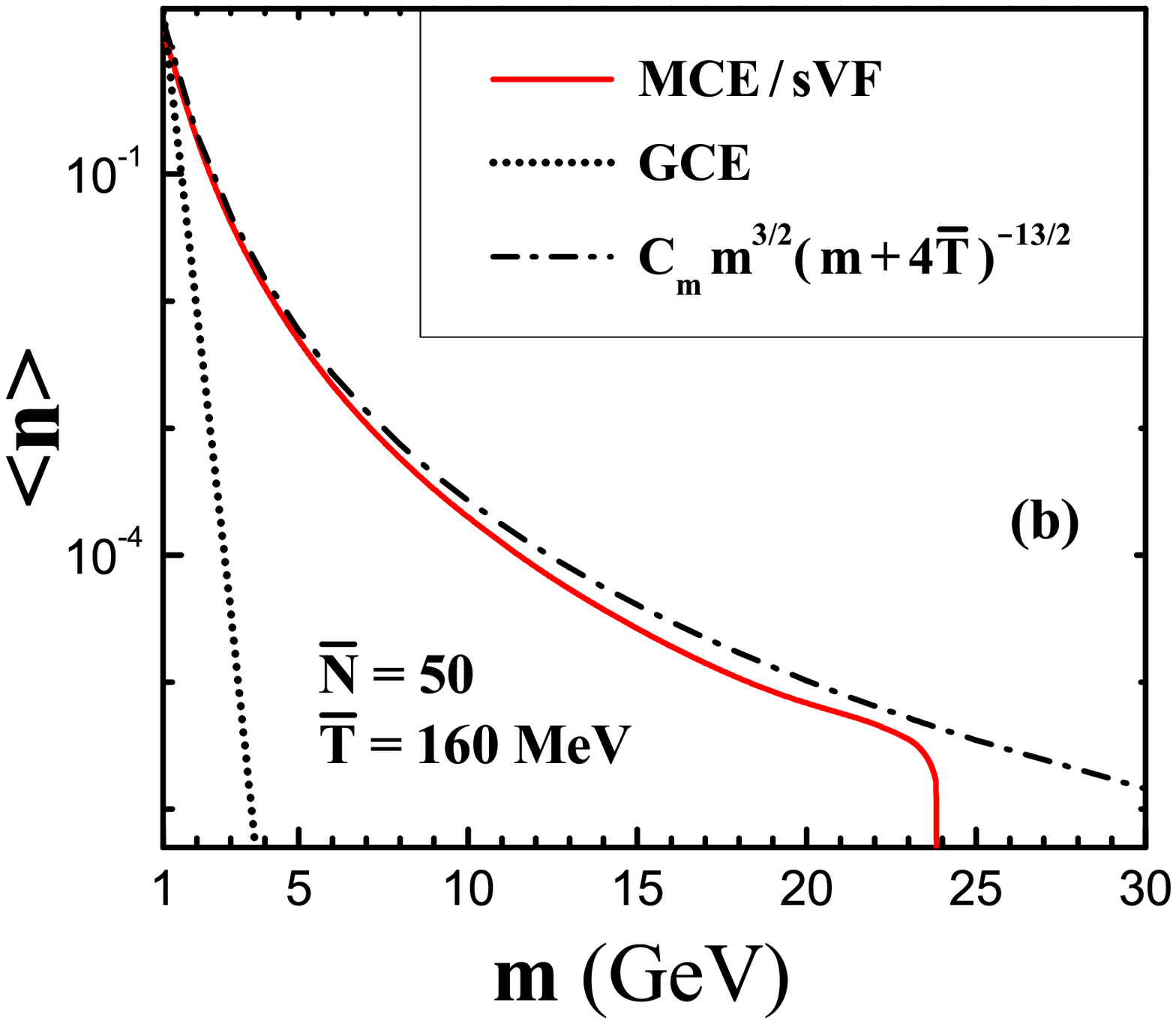,width=0.49\textwidth}
\caption{(Color online) {\bf (a):} The mean multiplicity of
neutral heavy particles, $\langle n\rangle$,  as a function of
heavy particle mass, $m$. Note, that the horizontal scale starts
at $m = 1$~GeV. The mass dependence obtained with the MCE
(\ref{n-mce-av}) is indicated by the dashed line, whereas the GCE
dependence (\ref{nM-GCE}) is shown by the dotted line. The system
energy is $E=3\overline{N}\,T=24$ GeV. {\bf (b):} The mass
dependence (\ref{n-a-av}) within the MCE/sVF is shown by the solid
line, and the dependence for the GCE (\ref{nM-GCE}) is indicated
by the dotted line. The approximation (\ref{ana3}) of the MCE/sVF
results is shown by the dashed-dotted line. The system energy is
$E=3\overline{N}\,\overline{T}=24$ GeV.} \label{fig-nM}
\end{figure}

The MCE partition function for $n$ heavy particles is given by
\cite{Fermi,mce2}:
 \eq{ \label{Wn(E,V)}
 \Omega_n(E,\,V )
 &\,\equiv\, \frac{1}{n!} \left(\frac{GV}{2\pi^2}\right)^n
    \int_0^{\infty} p_1^2 dp_1 \ldots \int_0^{\infty}p_n^2 dp_n\,
    \delta\left[E \,-\, \sum_{j=1}^{n}
    \left(m\,+\,\frac{p_j^2}{2m}\right)\right]\nonumber  \\
 &\,=\, \frac{1}{n!}\left(\frac{GV}{(2\pi)^{3/2}}\right)^n
        \frac{m^{\frac{3n}{2}}}{\Gamma\left(\frac{3n}{2}\right)}
        \left(E -n~m\right)^{\frac{3n}{2}-1}\,,
 }
where $m$ is the mass of the heavy particle and $G$ is its
degeneracy factor. Using Eqs.~(\ref{W(N)}) and (\ref{Wn(E,V)}) one
can calculate the MCE partition function for $N$ massless and $n$
heavy particles. It reads \cite{Fermi,mce2}:
 \eq{ \label{ZNn}
 Z_{N,n}(E,\,V)
 &\,=\, \int_0^{\infty}dE_1 \int_0^{\infty}dE_2\,
        W_N(E_1,\,V)\,\Omega_n(E_2,V) \delta[E-E_1-E_2] \nonumber \\
 &\,=\, \frac{1}{N!}\, \frac{1}{n!} \left(\frac{gV}{\pi^2}\right)^N
        \left(\frac{GV}{(2\pi)^{3/2}}\right)^n
        \frac{m^{\frac{3n}{2}}}
    {\Gamma(3N+\frac{3}{2}\,n)}\,(E -n~m)^{3N+\frac{3n}{2}-1}\,.
        }
In what follows the values of the degeneracy factors are
assumed to be  $g=G=1$.
Note, that due to the exact energy conservation, the MCE partition
functions, $W_N(E,\,V)$ (\ref{W(N)}) and $\Omega_n(E,V)$
(\ref{Wn(E,V)}), are defined for a particle number larger than zero,
$N\geq 1$ and $n\geq 1$.
However, the MCE partition function, $Z_{N,n}(E,\,V)$ (\ref{ZNn}),
requires only $N+n\geq 1$, so that either $N$ or $n$ can be equal
to zero.
The mean heavy particle multiplicity in the system
with two types of particles,
massless and heavy, is:
 \eq{ \label{n-mce-av}
 \langle n \rangle_{mce}
 ~=~ \frac{1}{Z(E,\,V)}
 \sum_{N,n=0}^{\infty} n~ Z_{N,n}(E,\,V)\,,
 }
where $Z_{N,n}(E,\,V)$ is given by Eq.~(\ref{ZNn}),
$Z(E,V)\equiv\sum_{N,n}^{\infty} Z_{N,n}(E,V)$, and
$Z_{0,0}(E,V)=0$. The energy conservation implies that there is a
maximum number of heavy particles, $n_{max}=
\left[\frac{E}{m}\right]$. Figure~\ref{fig-nM}a shows the
dependence of the mean multiplicity of heavy particles on mass $m$
calculated within the MCE (\ref{n-mce-av}). The plot starts at $m
= 1$~GeV. This is because at lower masses, $m \cong T=160$ MeV,
and thus the non-relativistic approximation used for heavy
particles is not justified. Furthermore, if $m\le 1$~GeV, a
contribution of heavy particles to the system energy density can
not be neglected, and this would change the temperature parameter
$T$ \cite{mce2}. An exponential decrease with the inverse slope
parameter of about 160~MeV observed for masses of several GeV, is
followed by a steepening decrease towards the threshold.

At low masses, $m\ll E$, the MCE mean multiplicity is close to the
corresponding multiplicity calculated within the GCE, see
Fig.~\ref{fig-nM}~(a):
\eq{\label{nM-GCE}
\overline{n}~=~V~(m\,T/2\pi)^{3/2}~
               \exp\left(-~\frac{m}{T}\right)~.
}
This can be shown analytically as follows. For the considered
values of the total energy $E$ and volume $V$ the average
multiplicity $\langle n\rangle_{mce}$ is essentially smaller than
one. Thus, the main contribution to the numerator of
Eq.~(\ref{n-mce-av}) comes from $Z_{N,n=1}$, while the main
contribution to the denominator comes from $Z_{N,n=0}$, thus:
 \eq{\label{nmce}
 \langle n\rangle_{mce}
 \;\cong\; \frac{\sum_{N=0}^{\infty}Z_{N,1}}{\sum_{N=1}^{\infty}Z_{N,0}}
 \;=\; \frac{\sqrt{2}\,m^{3/2}\sqrt{E-m}}{E^2}\;
       \frac{_0F_3\left(;\frac{1}{2},\frac{5}{6},\frac{7}{6},
       \frac{V(E-m)^3}{27\pi^2}\right)}
       {_0F_3\left(;\frac{4}{3},\frac{5}{3},2,
       \frac{V\,E^3}{27\pi^2}\right)}\;.
 }
Using Eq.~(\ref{F03}) from Appendix one can show that in the
region $T\ll m\ll E$ it follows:
 \eq{
 \langle n\rangle_{mce}\;\cong\; \overline{n}\;.
 }
Note that  Eq.~(\ref{nmce}) is valid not only for $\langle
n\rangle_{mce}\ll 1$, but for all values of $\langle
n\rangle_{mce}$ including also $\langle n\rangle_{mce}\gg 1$. In
the limit $V E^3\rightarrow\infty$ the distribution of heavy
particles in the MCE has the Poisson form $P_{mce}(n)\cong
\exp(-\overline{n})~\overline{n}^n/(n!)$, for $n\,m\ll E$. Thus, $
\langle n\rangle_{mce}\cong P_{mce}(1)/P_{mce}(0)\cong
\overline{n}$ at all $\overline{n}\ll E/m$.

The mean multiplicity of heavy particles within the MCE/sVF is
given by:
 \eq{\label{n-a-av}
 \langle n \rangle_{\alpha}
 \;=\; \int_0^{\infty}dy\;\psi_{\alpha}(y)\, \langle n\rangle_{mce}\;,
 }
and its dependence on the mass $m$ is shown in Fig.~\ref{fig-nM}
(b). This dependence resembles the behavior of the MCE/sVF
momentum spectrum for massless particles shown in
Fig.~\ref{fig-momentum}.

The mean multiplicity $\langle n \rangle_{\alpha}$ as a function
of $m$ obeys the power law behavior:
\eq{\label{Km}
\langle n\rangle_{\alpha}~\cong~ \overline{N}~C_m~m^{-K_m}~,
}
in the large central interval of heavy particle masses.
%Similar behavior is suggested by the experimental
%data~\cite{powerlaw}.
At high $m$, the $\langle n\rangle_{\alpha}$ is larger than
corresponding MCE and GCE values by many orders of magnitude. This
shows non-equivalence of the MCE/sVF and the GCE or the MCE for
mean multiplicity of heavy particles even in the thermodynamic
limit. Another example of such non-equivalence was already
obtained for the mean volume, see Eq.~(\ref{V-alpha}). For the
masses close to the threshold mass, $m = 24$~GeV, the mean
multiplicity steeply decreases to zero.
The power law approximation of the mass dependence can be derived
analytically:
% by an integration of the approximation (\ref{ana1})
%of the momentum spectrum for $m = M \gg T$, namely:
%
 \eq{\label{ana3}
& \langle n\rangle_{\alpha} \;=\;
\int_0^{\infty}dy\psi_{\alpha}(y)~\langle n\rangle_{mce} ~\cong~
%\left(\frac{m}{2\pi}\right)^{3/2}
%\int_0^{\infty}dVP_{\alpha}(V)~V~T^{3/2}~\exp\left(-\frac{m}{T}\right)\nonumber\\
\left(\frac{m\overline{T}}{2\pi}\right)^{3/2}\overline{V}
\int_0^{\infty}dy~\psi_{\alpha}(y)~
~y^{5/2}~\exp\left(-\frac{m~y}{\overline{T}}\right)~\nonumber
  \\
 &\;=\; \overline{N}~\frac{\sqrt{\pi}~k^k~\Gamma(k+5/2)}{2\sqrt{2}~\Gamma(k)}\;
         \frac{\overline{T}^{k+1}~m^{3/2}}{(m\,+\,\overline{T}k)^{k+5/2}}
  \;\cong\;\overline{N} \cdot 0.81 \cdot{\rm GeV}^5~
  \frac{m^{3/2}}{(m\,+\,\overline{T}k)^{13/2}}~.
  %\nonumber
 }
In the last expression $\overline{T}=160$~MeV and $k = 4$ are set.
The approximation in Eq.~(\ref{ana3}) is valid for the particle
masses much smaller than the system energy, $m \ll E$.  The
analytical expression (\ref{ana3}) describes the numerical results
on particle multiplicity at  $1$~GeV$<m\ll E$. This is illustrated
in Fig.~\ref{fig-nM}b, where  both are compared. If $m \gg
\overline{T}$, the power law dependence (\ref{Km}) of the mean
multiplicity on $m$ appears with $C_m\cong 0.81~\rm{GeV}^5$ and
$K_m=5$. This approximation is valid when there is a distinct
region of $m$ in which $\overline{T} \ll m \ll E$.

The relation $K_m=K_p-3$, suggested by the experimental data and
derived using a simple statistical considerations \cite{powerlaw},
is valid within the MCE/sVF. Note, that also the values of the
power law parameters, $K_p= 8$ and $K_m= 5$, are close to the
corresponding parameters extracted from the experimental data
\cite{powerlaw}. Within the MCE/sVF these parameters are, however,
dependent on the choice of the volume scaling function. For the
function (\ref{psi-b}) one gets $K_p\cong 5.7$ and $K_m\cong 2.5$.
Thus, a quantitative comparison with the data requires further
studies.

\section{Summary and Closing Remarks}

In this paper the statistical  approach to particle production in
high energy collisions is extended. A model called the
Micro-Canonical Ensemble with scaling Volume Fluctuations, or the
MCE/sVF, is proposed which incorporates volume fluctuations in the
micro-canonical ensemble. A class of scaling volume fluctuations
is considered. Analytical and numerical calculations are performed
within the simplest formulation of the model which preserves its
qualitative features.

First, it is  shown that the model leads to KNO scaling of the
multiplicity distribution. The volume scaling function is selected
in order to reproduce the KNO scaling function  measured in p+p
interactions.

Second, the single particle momentum spectrum is calculated for
massless particles. The MCE/sVF spectrum has features which
resemble properties of the transverse momentum spectrum of hadrons
produced in very high energy collisions. In particular, a long
power law tail appears as a result of the volume fluctuations.

Third, the dependence of the mean multiplicity of heavy particles
on the particle mass is obtained. The mass dependence
approximately obeys a power law decrease, similar to what is
observed experimentally. The power parameters for the momentum
spectrum and the mass dependence of mean multiplicity are related
as $K_m \cong K_p - 3$, again, similar to experimental data.

Fourth, it is also shown that for particles with arbitrary mass
the dependence of the momentum spectrum on the particle mass and
momentum reduces to the dependence on the particle energy.

%\vspace{0.7cm}
%It is, thus, not clear how to interpret the results of the
%statistical MCE/sVF model in which the power law tail appears
%as a result of {\bf volume fluctuations with a specific scaling
%assumption which leads to} the proper description of the KNO
%multiplicity scaling.

%The MCE/sVF model shows a remarkable property. When imposing
%the KNO scaling, the $p_T$-spectra come out with the correct power
%law behavior. This unexpected feature of the statistical model
%looks rather interesting and deserves further studies.
A quantitative comparison with the experimental data requires a
significant additional work, in particular the introduction of the
proper degrees of freedom and all related conservation laws.

For more than 30 years the production of hadrons at high
transverse momenta and/or with high masses, which follows the
power law dependence, is discussed within dynamical QCD based
models of parton scattering with a large momentum transfer. In
particular, the perturbative QCD calculations quantitatively
describe many experimental results. Thus, in view of the evident
success of the QCD, the applicability range of the MCE/sVF and a
relation of the model to the QCD based approaches should be
critically studied.
%Moreover, the QCD based approaches may suggest
%an interpretation of the scaling volume fluctuations.

Clearly, further tests of the MCE/sVF are needed. Other
qualitative features of  p+p interactions should be discussed. The
most important is the jet-like correlations of high transverse
momentum  particles. These correlations are usually described
within the perturbative QCD for hard parton scattering accompanied
with phenomenological parton-hadron fragmentation function. It is
not yet clear whether the jet structure of particle production at
high transverse momenta can be reproduced within the MCE/sVF.
Clearly this would require an introduction of the momentum
conservation. The multiplicity distributions in p+p collisions
obey some additional regularities,  e.g., the Golokhvastov scaling
of the semi-inclusive spectra~\cite{semi}. They  should be also
studied within the MCE/sVF.

An effort to extend the model to nucleus-nucleus collisions is
needed. First predictions can, however, be obtained from
Eqs.~(\ref{ana1}) and (\ref{ana3}). Namely, the particle momentum
distribution at high (transverse) momenta and the mean
multiplicity of heavy particles are expected to behave as:
\eq{\label{NpAA}
\frac{dN}{p^2dp}~&\cong ~C_p~\langle N\rangle~p^{-K_p}~,\\
\langle n \rangle ~&\cong C_m~\langle N\rangle~m^{-K_m}~,
\label{nMAA}
}
where parameters $K_p$ and $K_m$ in Eqs.~(\ref{NpAA}, \ref{nMAA})
depend only on the form of the volume scaling fluctuations, and
the parameters $C_p$ and $C_m$ depend, in addition, on the mean
temperature. Under the assumption that the volume scaling function
and mean temperature are energy and system size independent, the
MCE/sVF predicts the proportionality of $dN/(p^2dp)$ at large $p$
and of $\langle n\rangle$ at large $m$ to the average total
multiplicity, $\langle N\rangle$. Data on system size dependence
of the mean multiplicity of $J/\psi$ mesons as well as the
particle yield at high transverse momenta seem to confirm this
prediction~\cite{MG2,phobos}.

\begin{acknowledgments}
We would like to thank F.~Becattini, A.~S.~Golokhvastov,
W.~Greiner, M.~Hauer, I.~N.~Mishustin, St.~Mrowczynski and
P.~Seyboth for numerous discussions. This work was in part
supported by the Program of Fundamental Researches of the
Department of Physics and Astronomy of NAS, Ukraine and the
Institute VI~-~146 of Helmholtz Gemeinschaft, Germany. V.V.B.
would like also to thank for the support of The International
Association for the Promotion of Cooperation with Scientists from
the New Independent states of the Former Soviet Union (INTAS),
Ref. Nr. 06-1000014-6454.
\end{acknowledgments}

\appendix

\section{}

The generalized hyper-geometric function, also known as the Barnes
extended hyper geometric functions defined by the following series
\cite{I}:
\begin{equation}\label{GHF}
{}_p F_q(a_1,a_2,\dots,a_p;b_1,b_2,\dots,b_q;z)=
\sum_{k=0}^{\infty} \frac{(a_1)_k (a_2)_k \dots (a_p)_k~z^k}
{(b_1)_k (b_2)_k \dots (b_q)_k~k!} ~,
\end{equation}
where $(a)_k$ is the Pochhammer symbol:
\begin{equation}\label{Poch}
(a)_k \equiv \frac{\Gamma(a+k)}{\Gamma(a)} = a (a+1) \dots
(a+k-1)~.
\end{equation}
The Euler gamma function $\Gamma(x)$ has a simple form for integer
$k$ and half­integer $k+1/2$ arguments:
\eq{ \Gamma(k)\,=\,(k - 1)!\,,\quad
\Gamma\left(k+\frac{1}{2}\right)
  \,=\, \frac{1\cdot3\cdot \ldots \cdot (2k - 1)}{2^n}\, \sqrt{\pi}\,.
}
The asymptotic behavior at $z \rightarrow \infty$ of the
hyper-geometric function $_pF_q$ for p=0 and q=3 is given by
\cite{I}:
 \eq{\label{F03}
 _0F_3\left(;\,b_1,b_2,b_3;\,z\right)
  \;\cong\; \frac{\Gamma(b_1)\,\Gamma(b_2)\,\Gamma(b_3)}{4\sqrt{2}\,\pi^{3/2}}\;
  \exp\left[4\sqrt[4]{z}\right]\cdot
  z^{\frac{1}{4}\left(\frac{3}{2}\;-\;b_1\;-\;b_2\;-\;b_3\right)}
  \left(1\;+\;{\it }O\left(\frac{1}{\sqrt[4]{z}}\right)\right).
 }

%\begin{eqnarray}
%{}_0 F_q(;b_1,b_2,\dots,b_q;z) &\cong& \frac{\prod_{j=1}^{q}
%\Gamma(b_j)}{\sqrt{(q+1) (2 \pi)^q}}
%(\sqrt[q+1]{z})^{\frac{q}{2}-\sum_{j=1}^{q} b_j}
%e^{(q+1) \sqrt[q+1]{z}}
%\label{GHFas}
%\end{eqnarray}


\begin{thebibliography}{100}

\bibitem{Ha:65}
R. Hagedorn, Suppl. Nuovo Cimento {\bf 3}, 147 (1965).

\bibitem{Be:97}
F. Becattini and U. Heinz, Z. Phys. C {\bf 76},  269 (1997);
F.~Becattini, L.~Bellucci and G.~Passaleva, Nucl. Phys. {\bf B}
Proc. Suppl. {\bf 92}, 137 (2001).

\bibitem{kno}
Z. Koba, H. B. Nielsen, P. Olesen, Nucl. Phys. B {\bf 40},
317 (1972).

\bibitem{knog1}
P. Slattery, Phys. Rev. Lett. {\bf 29}, 1624 (1972);
Phys. Rev. D {\bf 7}, 2073 (1973).

\bibitem{knog2}
M. Gazdzicki, R. Szwed, G. Wrochna, and A. K. Wroblewski,  Mod.
Phys. Lett. A {\bf 6}, 981 (1991).

\bibitem{knog3}
A. I. Golokhavastov, Yad. Fiz. {\bf 64}, 88 (2001), {\it ibid}
{\bf 64}, 1924 (2001).

\bibitem{Be:04}
%\cite{Begun:2004gs}
%\bibitem{Begun:2004gs}
  V.~V.~Begun, M.~Gazdzicki, M.~I.~Gorenstein and O.~S.~Zozulya,
  %``Particle Number Fluctuations in Canonical Ensemble,''
  Phys.\ Rev.\  C {\bf 70}, 034901 (2004).
  %[arXiv:nucl-th/0404056].
  %%CITATION = PHRVA,C70,034901;%%

\bibitem{mce1}
 V. V.~Begun, M. I.~Gorenstein, A. P.~Kostyuk,
 and O.S.~Zozulya, Phys. Rev. C
 {\bf 71}, 054904 (2005).

\bibitem{Be:07}
 V.~V.~Begun, M.~I.~Gorenstein, M.~Hauer, V.~P.~Konchakovski and O.~S.~Zozulya,
  %``Multiplicity fluctuations in hadron-resonance gas,''
 Phys. Rev. C {\bf 74}, 044903 (2006);
   V.~V.~Begun, M.~Gazdzicki, M.~I.~Gorenstein, M.~Hauer, V.~P.~Konchakovski and B.~Lungwitz,
  %``Multiplicity fluctuations in relativistic nuclear collisions:   statistical
  %model versus experimental data,''
  Phys.\ Rev.\  C {\bf 76}, 024902 (2007);
 M.~Hauer, V. V.~Begun, and M. I.~Gorenstein, arXiv:0706.3290 [nucl-th].
 %\cite{Begun:2006uu}

\bibitem{powerlaw}
%\cite{Gazdzicki:2001da}
%\bibitem{Gazdzicki:2001da}
  M.~Gazdzicki and M.~I.~Gorenstein,
  %``Power law in hadron production,''
  Phys.\ Lett.\  B {\bf 517}, 250 (2001).
  % [arXiv:hep-ph/0103010].
  %%CITATION = PHLTA,B517,250;%%

\bibitem{alpha}
 M. I. Gorenstein and M. Hauer, arXiv:0801.4219 [nucl-th].



\bibitem{pressure}
M. I. Gorenstein, Sov. J. Nucl. Phys. {\bf 31}, 845 (1980);
M.~I.~Gorenstein,
%``Isobaric Ensemble In The Bag Model And Multiplicity Distribution. (In
%Russian),''
Yad.\ Fiz.\  {\bf 31} 1630 (1980). %(In Russian).

\bibitem{stas}
%\cite{Mrowczynski:1984tj}
%\bibitem{Mrowczynski:1984tj}
  S.~Mrowczynski,
  %``On Multiplicity Distributions And A Pressure Ensemble,''
  Z.\ Phys.\  C {\bf 27}, 131 (1985).
  %%CITATION = ZEPYA,C27,131;%%

 \bibitem{Fermi}
 E.~Fermi,
  %``High-energy nuclear events,''
  Prog. Theor. Phys.  {\bf 5}, 570 (1950).

\bibitem{pp_spectra}
E. W. Beier {\it et al.}, Phys. Rev. D {\bf 18}, 2235 (1978).

\bibitem{wilk}
%\cite{Wilk:1999dr}
%\bibitem{Wilk:1999dr}
  G.~Wilk and Z.~Wlodarczyk,
  %``On the interpretation of nonextensive parameter q in Tsallis statistics
  %and Levy distributions,''
  Phys.\ Rev.\ Lett.\  {\bf 84}, 2770 (2000).
  %[arXiv:hep-ph/9908459].
  %%CITATION = PRLTA,84,2770;%%

\bibitem{tsallis}
C. Tsallis, J. Stat. Phys. {\bf 52}, 479 (1988).

\bibitem{Wilk:2006vp}
  G.~Wilk and Z.~Wlodarczyk,
  %``Fluctuations, correlations and the nonextensivity,''
  Physica A {\bf 376}, 279 (2007)
%  [arXiv:cond-mat/0603157].
  %%CITATION = PHYSA,A376,279;%%

\bibitem{mce2}
 V. V. Begun, L. Ferroni, M. I. Gorenstein, M. Gazdzicki,
 F.~Becattini,
% Threshold effects in relativistic gases,
 J. Phys. G {\bf 32}, 1003 (2006).


\bibitem{semi}
%\cite{Golokhvastov:2002cf}
%\bibitem{Golokhvastov:2002cf}
  A.~I.~Golokhvastov,
  %``Scaling of semi-inclusive events in p p interactions. (In Russian),''
  Phys.\ Atom.\ Nucl.\  {\bf 67}, 337 (2004)
  [Yad.\ Fiz.\  {\bf 67}, 355 (2004)].
  %%CITATION = YAFIA,67,355;%%

\bibitem{MG2}
%\cite{Gazdzicki:1999rk}
%\bibitem{Gazdzicki:1999rk}
  M.~Gazdzicki and M.~I.~Gorenstein,
  %``Evidence for statistical production of J/psi mesons in nuclear  collisions
  %at the CERN SPS,''
  Phys.\ Rev.\ Lett.\  {\bf 83}, 4009 (1999).
  % [arXiv:hep-ph/9905515].
  %%CITATION = PRLTA,83,4009;%%

\bibitem{phobos}
%\cite{Back:2003qr}
%\bibitem{Back:2003qr}
  B.~B.~Back {\it et al.}  [PHOBOS Collaboration],
  %``Charged hadron transverse momentum distributions in Au + Au collisions  at
  %s(NN)**(1/2) = 200-GeV,''
  Phys.\ Lett.\  B {\bf 578}, 297 (2004).
  % [arXiv:nucl-ex/0302015].
  %%CITATION = PHLTA,B578,297;%%


\bibitem{I}
Weisstein,~Eric~W.~"Generalized Hypergeometric Function." From
{\it MathWorld} -- A Wolfram Web Resource.
http://mathworld.wolfram.com/GeneralizedHypergeometricFunction.html,
http://functions.wolfram.com/HypergeometricFunctions/HypergeometricPFQ/06/02/05/
%M. Abramowitz and I. E. Stegun,
% {\it Handbook of Mathematical Functions}
%(Dover, New York, 1964).



\end{thebibliography}
\end{document}